\documentclass[review]{elsarticle}
\usepackage{epstopdf}
\usepackage{lineno,hyperref}
\modulolinenumbers[5]
\renewcommand{\figurename}{Fig.}
\renewcommand{\tablename}{Tab.}
\usepackage{booktabs}
\usepackage{longtable}
\usepackage{color}
\usepackage{verbatim}

\journal{Journal of Fluids $\&$ Structures Templates }

\begin{document}

\begin{frontmatter}

\title{Formulations of Hydrodynamic Force in the Transition Stage of the Water Entry of Linear Wedges with Constant and Varying Speeds}
% A new title should be rewritten to clarify the new contribution of the present paper from the recent paper, the abstract and conclusion also need some changes.
\author[mymainaddress]{Xueliang Wen}
\ead{wenxueliang@buaa.edu.cn}

\author[mymainaddress]{Peiqing Liu}
\ead{lpq@buaa.edu.cn}

\author[mysecondaryaddress]{Alessandro Del Buono}
\ead{alessandro.delbuono@inm.cnr.it}

\author[mymainaddress]{Qiulin Qu}
\cortext[mycorrespondingauthor]{Qiulin Qu}
\ead{qql@buaa.edu.cn}

\author[mysecondaryaddress]{Alessandro Iafrati}
\ead{alessandro.iafrati@cnr.it}

\address[mymainaddress]{School of Aeronautic Science and Engineering, Beihang University, Beijing, 100191, China}

\address[mysecondaryaddress]{CNR-INM, Via di Vallerano 139, 00128 Rome, Italy}

\begin{abstract}
There is an increasing need to develop a two-dimensional (2D) water entry model including the slamming and transition stages for the 2.5-dimensional (2.5D) method being used on the take-off and water landing of seaplanes, and for the strip theory or 2D+t theory being used on the hull slamming. Motivated by that, this paper numerically studies the transition stage of the water entry of a linear wedge with constant and varying speeds, with assumptions that the fluid is incompressible, inviscid and with negligible effects of gravity and surface tension, and the flow is irrotational. For the constant speed impact, the similitude of the declining forces of different deadrise angles in the transition stage are found by scaling the difference between the maximum values in the slamming stage and the results of steady supercavitating flow. The formulation of the hydrodynamic force is conducted based on the similitude of the declining forces in the transition stage together with the linear increasing results in the slamming stage. For the varying speed impact, the hydrodynamic force caused by the acceleration effect in the transition stage is formulated by an added mass coefficient with an averaged increase of $27.13\%$ compared with that of slamming stage. Finally, a general expression of the hydrodynamic forces in both the slamming and transition stages is thus proposed and has good predictions in the ranges of deadrise angles from $5^\circ$ to  $70^\circ$ for both the constant and varying speed impacts.
\end{abstract}


\end{frontmatter}

\section*{Nomenclature}
\begin{longtable}{cll}
$\bf a$&=&Accelerations of body\\
$a_w$&=&Acceleration of body in vertical direction\\
$A$&=& Dimensionless variable of the theory of steady supercavitating flow \\
$A_2$, $B_2$&=& Dimensionless variables of approximate solution \\
$A_{2\rm Mei}$, $B_{2 \rm Mei}$&=& Dimensionless variable of Mei et al.'s model\\
$c$&=&Effective wetted length\\
$C_a$&=&Added mass coefficient\\
$C_{a0}$&=& $A_2(k_2\tan\beta)^2$ Added mass coefficient in the slamming stage\\
$C_{\rm const}$&=& Dimensionless coefficient of constant speed impact \\
$C_{\rm Korobkin}$&=& $C_{\rm const}$ of Korobkin's model \\
$C_p$&=&Pressure coefficient\\
$C_{p0}$&=&Pressure coefficient  of constant speed impact\\
$c_q$&=&Volume fraction of $q^{th}$ fluid\\
$C_{s}$&=&Slamming coefficient\\
$C_{s0}$&=&Slamming coefficient of constant speed impact\\
$C_{s0}^*$&=&$({C_{s0}-C_{s\infty}})/({C_{s\rm max}-C_{s\infty}})$\\
$C_{s\rm max}$&=&Maximum Slamming coefficient\\
$C_{s\infty}$&=&Slamming coefficient of steady supercavitating flow\\
$f_{\rm 3D}$&=&Force coefficient of 3D effect\\
$F$&=&Force acting on bodies in vertical direction\\
$\mathbf{F}$&=&Total hydrodynamic forces acting on the body surface\\
$F_0$&=&Force acting on bodies of constant speed impact\\
$Fr$&=&$U_0/\sqrt{{\rm g}l}$ Froude number of the freefall motion case\\
$\rm g$&=& Gravitational acceleration \\
$\mathbf g$&=&  Body forces in Euler equations\\
$G$&=&Core function of the second Green identity\\
$h$&=&Penetration depth\\
$h_0$&=& $l\tan\beta$ Height of wedge\\
$h_2$&=& Penetration depth corresponding to $C_{s\rm max}$\\
$h^*$&=& $({h-h_2})/{h_2}\cot^\lambda\beta$\\
$k$&=&Parameter of curved FBC\\
$k_1$&=&$h_2/h_0$ Dimensionless penetration depth of the maximum $C_{s0}$\\
$k_2$&=&$\frac{\partial C_{s0}}{\partial (h/h_0)}$ Dimensionless derivative of $C_{s0}$ with respect to $h$\\
$l$&=&Half width of a finite body\\
$L$&=&Thickness of a 3D wedge\\
$m$&=&Mass of the wedge\\
$m_0$&=&Added mass of wedge in half model\\
$\bf n$&=&Unit vector of normal to the wall surface\\
$ n_y$&=&Projection in vertical direction $y$ of $\bf n$ \\
$N$&=&Number of nodes to calculate parameter $n$ \\
$p$&=&Pressure\\
$p_0$&=&Pressure of constant speed impact\\
$p_a$&=&Atmosphere pressure\\
$\bar P$&=& Vertices of the lower side in jet region\\
$\bar P$&=& Vertices of the upper side in jet region\\
$r$&=& Ratio between the normal distance between the two side\\
 &&of the jet in proximity of the spray root and the minimun panel size\\
$t$&=&Time\\
$\mathbf{u}$&=&$\nabla \phi$ Velocities of fluid\\
$u_n$, $u_s$&=&Normal and tangential projections of fluid velocity on the body surface\\
$U_0$&=&Initial speed of wedge\\
$U_w$&=&Instantaneous speed of wedge\\
$v$&=&Fluid velocity of the theory of steady supercavitating flow\\
$v_0$&=&Incoming velocity of water  of steady supercavitating flow\\
${\bf V}_q$&=&Velocities of the $q^{th}$ fluid \\
$v_n$&=&Vertical velocity on the free surface\\
$\bf V$&=&Velocities of fluid\\
$w_n$, $w_s$&=&Normal and tangential projections of body speeds on the body surface\\
$\bf W$&=&Body speeds\\
$\bf x$&=&Position of particles lying on the free surface\\
$x$, $y$,&=&Cartesian coordinates\\
$z$&=&$x+{\rm i}y$ Complex coordinates\\
$\alpha$, $\chi$, $\mu$, &=&Coefficients of the velocity potential expression in the jet region\\
$\beta$&=&Deadrise angle of wedge bodies\\
$\beta_L$&=&Deadrise angle of linear FBC\\
$\Delta p$&=&Difference of pressure between the constant and varying speed impacts \\
$\Delta F$&=&Difference of force between the constant and varying speed impacts \\
$\eta$&=&  $\frac{1}{N}\sum\limits_{i = 1}^N {\sigma _i^2}$ Mean $\sigma$ of all the nodes in $h^*\in [0,\,\,7]$ \\
$\gamma$&=&Correction factor of wetted length\\
$\gamma_{\rm Mei}$&=&Correction factor of wetted length for Mei et al.'s model\\
$\kappa$&=&Growth factor of BEM-FEM panels\\
$\lambda$&=&Auxiliary variable to model the $C_{s0}$ in the transition stage\\
$\omega$&=&Parameter complex plane \\
$\phi$&=&Velocity potential\\
$\phi_0$&=&Velocity potential at C and C' in the theory of steady supercavitating flow\\
$\Phi$&=&Dimensionless velocity potential of similarity solution\\
$\rho$&=&Density of water\\
$\sigma$&=&Standard deviation of $C_{s0}^*$ between different deadrise angles\\
$\tau$&=&Parameter complex plane \\
$\theta$&=&Argument of the velocity of fluid\\
$Superscript$\\
$J$&=& Panel in the jet region\\
$*$&=& Variables at the centroid of fluid control volume\\
$Subscript$\\
$q$&=& Phase of fluid\\
$i$&=& Index of vertices in the jet region or  index of node\\
\end{longtable}

\section{Introduction}\label{Part:1}

\begin{figure}
  \centerline{\includegraphics[width=12cm,height=5.2cm]{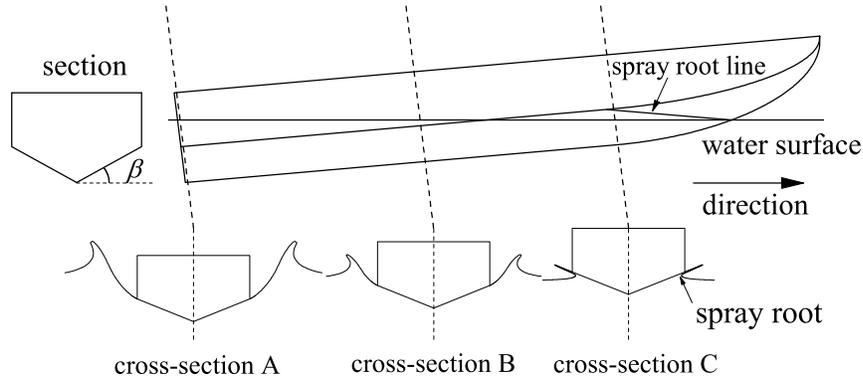}}
  \caption{The free surfaces of different cross-sections of a hull during the high speed planning.}
  \label{Fig:hull_sections}
\end{figure}

The take-off and water-landing of seaplanes had been studied since 1920s and the procedures to assess the structural crashworthiness of airframe is mainly based on a 2.5-dimensional (2.5D) method \cite{Mayo1945Analysis, Milwitzky1948A,smiley1951semiempirical}, which is similar to the formulations of slamming forces of the strip theory for the hull slamming. Every individual cross-section normal to the longitudinal direction of hull independently experiences a single process of water entry (see  \figurename$\,$\ref{Fig:hull_sections}), where the 2D water entry model is formulated by the added mass method \cite{von1929impact,Mayo1945Analysis, Milwitzky1948A} or the Wagner theory \cite{wagner1932stoss,smiley1951semiempirical}. The traditional procedures had been identified to be less accurate as the numerical methods have fast developed in recent decades, and make it possible to conduct numerical simulations of water-landing of seaplanes \cite{Numerical2018duan} and transport airplanes \cite{guo2013effect,qu2015study}. Although a whole-time history of water entry of aircraft can be reproduced, massive computational resources are required and several days of calculations are taken. This could not be treated as a practical method for the engineers to complete the initial designs of seaplanes or high-speed planning hulls \cite{savitsky1964Hydrodynamic,savitsky2007Inclusion}. The alternative method is to improve the 2D model of the 2.5D methods because the errors of early 2.5D methods are mainly resulted from the transverse pressure distribution, especially for that with chine immersion. For the cross-section C \figurename$\,$\ref{Fig:hull_sections} without chine immersion, they used the Wagner theory; For the cross-sections A and B in \figurename$\,$\ref{Fig:hull_sections} with chine immersion, they adopted the pressure distribution of a steady supercavitating flow \cite{korvin1948discontinuous,gurevich1965theory} to formulate the transverse flow. However, the pressure distribution of Wagner theory is quite different from that of steady supercavitating flow. How the pressure distribution continuously changes from that of Wager theory to that of steady supercavitating flow is missing in the early 2.5D methods. 
\begin{figure}
  \centerline{\includegraphics[width=12cm,height=4cm]{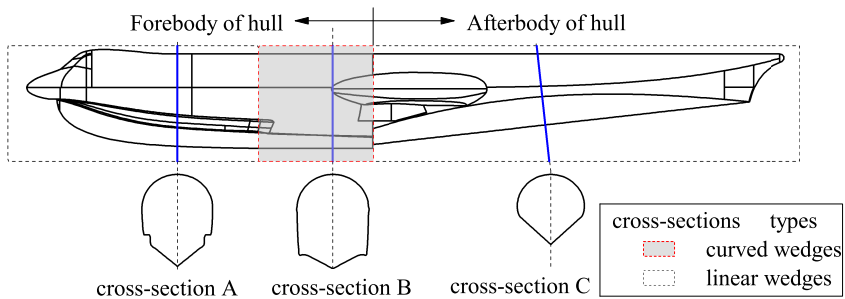}}
  \caption{The cross-sections (normal to the keels) of the hull of seaplanes.}
  \label{Fig:seaplane_description}
\end{figure}

\begin{figure}
  \centerline{\includegraphics[width=10cm,height=6cm]{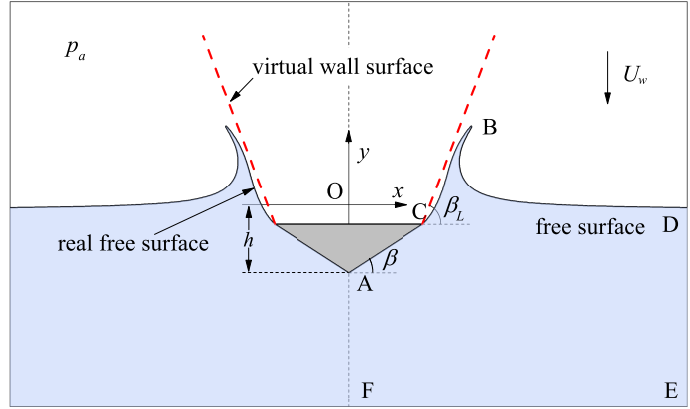}}
  \caption{The description of the water entry of a 2D wedge with a speed of $U_w$, $h$ being the penetration depth.}
  \label{Fig:transition_stage_description}
\end{figure}

From the perspective of a 2D transverse flow in a constant speed, the cross-section without chine immersion is corresponding to a slamming stage in which the hydrodynamic force increases linearly with the increasing wetted length, and the cross-section with chine immersion is corresponding to a transition stage in which the body experiences a fast drop of hydrodynamic force when the spray root leaves the chine. The pressure distribution and the hydrodynamic force gradually decline and finally approach those of steady supercavitating flow, as the experimental results of Zhao et al. \cite{zhao1996water} indicated. The poor accuracy of original 2.5D methods is due to the weakness of their 2D model for the transverse flow and lack of involvement of the hydrodynamic forces acting on afterbody of hull (see \figurename$\,$\ref{Fig:seaplane_description}). Since there is no effective theoretical method to formulate the transition stage, Sun and Faltinsen  \cite{Hui2007The,Hui2012Hydrodynamic} directly adopted a boundary element method (BEM) as the 2D water entry model to develop another similar method called 2D+t theory for the high speed hulls. Their BEM \cite{zhao1993water, zhao1996water} can address both the slamming and transition stages with increasing accuracy of predictions, but at the same time greatly increases the computational cost, which undermines the efficiency of the 2D+t theory. Therefore, there is an increasing need to improve the efficiency of the 2.5D methods and the 2D+t theory by proposing an analytical solution as accurate as the numerical methods to predict the hydrodynamic forces in both slamming and transition stages. For the slamming stage, many researchers had contributed to the formulations of hydrodynamic force by added mass methods  \cite{von1929impact,wagner1932stoss,faltinsen1993sea}, asymptotic theories \cite{logvinovich1969hydrodynamics,mei1999water,korobkin2004analytical} and approximate solution \cite{wen2020Numerical}. Among them, the approximate solution of Wen et al.\cite{wen2020Numerical} provided the most accurate model for the constant and varying speed cases by a similarity solution of  Dobrovol1'skaya\cite{dobrovol1969some}. For the transition stage, due to the different hydrodynamic characteristics, the formulation of hydrodynamic force is more difficult and is yet to be fully addressed. In this paper, by following the formulation work of our previous research \cite{wen2020Numerical},
 the transition stage of the water entry of linear wedges with constant and varying speeds, as shown in \figurename$\,$\ref{Fig:transition_stage_description}, is formulated to complete the 2D water entry model.

There are three different formulations of the hydrodynamic forces in the transition stage in literature. The first one was proposed by Logvinovich\cite{logvinovich1969hydrodynamics} based on the boundary condition at the separation (point C in \figurename$\,$\ref{Fig:transition_stage_description}). The pressure at C is required to be same with the atmospheric pressure, which is denoted as zero pressure condition. Logvinovich derived an ordinary differential equation (ODE) to obtain a virtual wetted length but integrated the pressure to obtain the hydrodynamic force on the real wetted surface AC. Tassin et al. \cite{Tassin2014On} improved the Logvinovich's model by introducing the correction of 1+$\tan^2\beta$ ($\beta$ being the deadrise angle) to the pressure expression on the wall surface and re-derived the ODE of virtual wetted length. The predictions are in better agreement with BEM results of Iafrati and Battistin \cite{Iafrati2003Hydroelastic} than the original one. The second way is called fictitious body continuation (FBC)  and was also developed by Tassin et al. \cite{Tassin2014On} based on a modified Logvinovich's model (MLM) of Korobkin \cite{korobkin2004analytical} who addressed the formulation of the slamming stage. The FBC is a virtual wall surface extended from the separation C with an angle of $\beta_L$ with respect to the horizonal line (see \figurename$\,$\ref{Fig:transition_stage_description}). The wetted length is solved from the combination of the real wall surface AC and FBC, while the hydrodynamic force is only integrated on AC as it does in the Logvinovich's model of the zero pressure condition. Tassin et al. \cite{Tassin2014On} had to compare with the numerical results to identify the parameter $\beta_L$. Although the agreement with the numerical results is better than the Logvinovich's model with correction 1+$\tan^2\beta$, there is still some discrepancy and the FBC needs more improvement. Wen et al. \cite{wen2022Modified} proposed the curved FBC to improve the accuracy of linear FBC of Tassin et al. \cite{Tassin2014On} based on a modified Wagner's model (MWM). They introduced another parameter $k$ and provided explicit equations to determine $\beta_L$ and $k$. Their predictions were in better agreement with the numerical results than the linear FBC of Tassin et al. \cite{Tassin2014On}. Since these methods are all based  on the Wagner theory \cite{wagner1932stoss}, their predictions will become less accurate when the deadrise angle is larger than 30$^\circ$. The last and the most sophisticated formulation was conducted by Semenov and Wu \cite{Semenov2016Water} by extending their integral hodograph method (IHM) \cite{semenov2006nonlinear, semenov2016liquid} from the slamming stage to transition stage. Their results show a larger declining force than the BEM simulations of Iafrati and Battistin \cite{Iafrati2003Hydroelastic} because they imposed a self-similar solution onto a non-self-similar flow in the transition stage. In general, the fully analytical solutions as accurate as the numerical results and with a range of deadrise angle $[15^\circ,\,\,45^\circ]$ for the applications of take-off and water landing of seaplanes are currently unavailable. In contrast to formulate the hydrodynamic force by the asymptotic analysis and self-similar solution, a semi-analytical solution for both the slamming and transition stages can be probably built by summarizing the numerical results and exploring the possible patterns of the hydrodynamic forces.

The failures of the abovementioned theoretical methods to formulate the hydrodynamic forces in the transition stage is caused by inaccurate modelling of free surface, which is the main difference between the numerical methods and the theoretical methods. There are two broad categories of the numerical methods: BEM and computational fluid dynamics (CFD). The free surface problems of water entry for the BEM include the modelling of jet and the new free surface AC. Zhao et al. \cite{zhao1996water} provided the first fully nonlinear BEM results, using a model of jet-cut to avoid the high-speed thin jet, and the lowest-order expansion of the Kutta condition (the flow leaves the body tangentially at the separation point C) to simplify the new free surface AC.  Iafrati and Battistin \cite{Iafrati2003Hydroelastic} proposed a different approach to switch boundary conditions (BCs) of the panels from a wall panel to a free surface panel when the center of panel leaves the separation point C. Instead of using the jet cutting strategy as Zhao and Faltinsen \cite{zhao1993water} and their previous method \cite{battistin2003hydrodynamic} did, they combined the use of a BEM solver in the bulk of fluid and a simplified finite element method (FEM) in the thin jet developing along the body contour. The hybrid method was denoted as a hybrid BEM-FEM (HBF) approach and can provide a detailed description of the flow and free surface dynamics, together with an improved prediction of the separation, while keeping the computational effort still reasonable. The HBF method was recently extended by Del Buono et al. \cite{del2021water} to deal with the water entry with varying speed and the water exit problems. Considering the good application of the BCs switching, Wang and Faltinsen \cite{Jingbo2013Numerical} also develop their BEM of jet cutting replacing the lowest-order expansion of the Kutta condition with the BCs switching for the water entry problems. In contrast to the BCs switching and FEM solver, Bao et al. \cite{Bao2017Simulation} used the least orders of equations to update the normal velocity of the free surface AC and a shallow water assumption for the jet region in the transition stage. The abovementioned BEM methods have different strategies for the jet region and new free surface AC, but are basically consistent with each other. They can reduce the computational cost and provide the velocity potential distribution on the wedge surface, which is important for the varying speed impact. Different from the complexity of the free surface modelling of BEM, the CFD methods are more flexible to deal with the modelling of free surface. The most widely used method for the free surface problems, volume of fraction (VOF), was first proposed by Hirt and Nichols \cite{hirt1981volume} based on the framework of  finite volume method (FVM).  A modified high-resolution interface capturing scheme (Modified HRIC) was developed by Muzaferija et al. \cite{muzaferija1999two} to solve volume fraction equations and has good applications to the water entry problems \cite{qu2015study,zhang2018numerical,wen2020Impact}. Compared with the BEM approaches, the CFD methods provide more information about the distributions of pressure and velocities in the whole region which intuitively picture the rapid changes of flow field in the transition stage of the constant speed impact. The FVM with VOF has been the most successful method to deal with the water entry problems and will be adopted in this paper to produce the numerical results of constant speed impact. In order to study the distribution of velocity potential on the wall surface for the varying speed impact, the HBF of Iafrati and Battistin \cite{Iafrati2003Hydroelastic} will also be used to provide the required data.

In this paper, the slamming and transition stages of water entry of linear wedges with constant and varying speeds are numerically studied by the FVM with VOF \cite{wen2020Impact} and the HBF of Iafrati and Battistin \cite{Iafrati2003Hydroelastic}. To propose a semi-analytical solution of a combination of numerical results and theoretical results to address the high speed impact problems, the fluid is considered to be incompressible, non-viscous, weightless and with negligible surface tension effects and the flow is to be irrotational. The hydrodynamic forces in the slamming and transition stages during the water entry of a wedge with constant and varying speeds will be formulated. The arrangements of this paper are as follows. The computational approaches of FVM with VOF and HBF are detailed in Sec.$\,$\ref{Part:2} and the validations of the two methods are given in Sec.$\,$\ref{Part:3}. For the constant speed impact, the formulation of forces in both slamming and transition stages of the water entry of linear wedges with the deadrise angles varying from $5^\circ$ to $70^\circ$ is proposed based on the CFD results in Sec.$\,$\ref{Part:4}. For the varying speed impact, the acceleration effect is addressed based on the HBF results in Sec.$\,$\ref{Part:5}. The general formulations of hydrodynamic force of both the slamming and transition stages will be provided and can address the constant speed impact and the acceleration effect.

\section{Computational Approaches}\label{Part:2}
A CFD method of FVM with VOF technique is adopted to provide the detailed results of flow field of water entry in a constant speed and the HBF  \cite{Iafrati2003Hydroelastic,battistin2004numerical} is used to calculate the velocity potential  (the CFD method can't provide the velocity potential) and the pressure distributions on the wedge surface during the water entry in a varying speed.
\subsection{CFD method}\label{Part:2.1}
\subsubsection{Flow solver}\label{Part:2.1.1}
The unsteady incompressible Euler equations ignoring the surface tension force are solved using ANSYS FLUENT as follows
\begin{equation}\label{Eq:57x}
\frac{{\partial{\bf V}}}{{\partial t}} + {\bf V} \cdot \nabla{\bf V} =  - \frac{1}{\rho }{\bf \nabla} p - \mathbf{g},
\end{equation}
where $\bf V$ is the velocity of fluid, $\rho$ is the density, $p$ is the pressure and $\mathbf{g} = (0, -\rm g)$ representing the gravity of fluid (the gravity of fluid can be neglected for the high speed impact). The semi-implicit method for pressure linked equation consistent algorithm (SIMPLEC) is used to deal with the pressure-velocity coupling. The unsteady terms are discretized by first order implicit scheme, the convention terms are discretized by second order upwind scheme, and the pressure term is discretized by body force weighted scheme.
\subsubsection{VOF method}\label{Part:2.1.2}
The VOF method was firstly proposed by Hirt and Nichols \cite{hirt1981volume}, which can capture the free interfaces between two or more immiscible fluids by introducing a variable, called volume fraction, for each phase. If the volume fraction of the $q^{th}$ fluid in a certain cell is denoted as $c_q$, $c_q=0$ represents the cell is empty of the $q^{th}$ fluid; $c_q=1$ represents the cell is full of the $q^{th}$ fluid; and $0<c_q<1$  represents the cell contains the interface between the $q^{th}$ fluid and other fluids. The sum of the volume fractions of all phases must be 1 in each cell. The volume fraction equation of the $q^{th}$ fluid is written as follows:
\begin{equation}\label{Eq:57}
\frac{\partial }{{\partial t}}({c_q}{\rho _q}) + \nabla  \cdot ({c_q}{\rho _q}{\bf V}_q) = 0,
\end{equation}
where ${\bf V}_q$ is the velocity of $q$ fluid. The first term in the left hand is discretized by one order implicit scheme, and the second term is discretized by modified high resolution interface capturing (Modified HRIC) scheme \cite{muzaferija1999two}.

\subsubsection{GMM method and VOF boundary conditions}\label{Part:2.1.3}
In this paper, the global moving mesh method (GMM) \cite{qu2015numerical} is used to deal with the motion of wedge. The whole computational domain (including the cells and boundaries) moves together with the wedge like a rigid body. The volume fraction boundary conditions can ensure that the free water surface keeps a given level when the computational domain moves. This condition is set according to the cell coordinates of the boundaries in the earth fixed coordinate system, i.e., the volume fraction of water $c_q=0.5$ for the cell located on the interface between air and water; $c_q=0$ for the cells located above the interface; $c_q=1$ for the cells located below the interface.

\subsection{HBF approach}\label{Part:2.2}
In this section, the fully nonlinear potential flow model based the hybrid BEM-FEM approach is presented. The method is mainly based on the studies of Refs.$\,$\cite{Iafrati2003Hydroelastic, battistin2004numerical}, where full details are provided.
\subsubsection{Governing Equations}
The water entry problem of rigid bodies is faced under the hypotheses of an inviscid and incompressible fluid. The flow is assumed irrotational and the problem is formulated in terms of the velocity potential $\phi$. Surface tension effects are neglected. The flow is therefore governed by the following initial-boundary value problem:
\begin{equation}\label{Eq:1}
\nabla^2 \phi = 0 \quad{\rm{in\,\,water\,\,domain}}\,\,\Omega,
\end{equation}
\begin{equation}\label{Eq:2}
\frac{\partial \phi } {\partial n}= {\bf W}\cdot{\bf n}=-U_wn_y\quad {\rm on \,\,wall \,\,surface\,\,}S_{\rm B},
\end{equation}
\begin{equation}\label{Eq:3}
\frac{{{\rm{D}}\phi }}{{{\rm{D}}t}} = \frac{{{{\left| {\nabla \phi } \right|}^2}}}{2}-{\rm g}y,\quad\frac{{{\rm{D}}{\bf{x}}}}{{{\rm{D}}t}} = \nabla \phi,\quad {\rm on\,\, free\,\, surface\,\,}S_{\rm S},
\end{equation}
where $\bf n$ is the unit vector of normal to the wall surface, $ n_y$ is the projection of $\bf n$ in the vertical direction $y$, ${\bf W}$ is the entering speed of the body, and $\bf x$ is the position of the particle lying on the free surface. At each time step, the solution of the boundary-value problem for the velocity potential is solved in the form of the boundary integral representation provided by the second Green’s identity
\begin{equation}\label{Eq:7}
\frac{{\rm{1}}}{{\rm{2}}}\phi \left( P \right) = \int_{{S_{\rm B}} \cup {S_{\rm S}}} {\left[ {{\phi _n}\left( Q \right)G\left( {P,Q} \right) - \phi \left( Q \right){G_n}\left( {P,Q} \right)} \right]{\rm{d}}S\left( Q \right)} ,\,\, P \in S_{\rm B} \cup S_{\rm S}
\end{equation}
where $G\left( {P,Q} \right) = \frac{1}{{2\pi }}\log \left( {\left| {P - Q} \right|} \right)$. According to Eqs.$\,$(\ref{Eq:2})-(\ref{Eq:3}), the velocity potential is known on the free surface while its normal derivative is assigned on the body contour, which belongs to a boundary integral equation of mixed first and second kind. Once Eq.$\,$(\ref{Eq:7}) is solved, the velocity potential and its normal derivative are known on the body contour and the free surface. The solution of the boundary integral equation Eq.$\,$(\ref{Eq:7}), providing the normal derivative of the velocity potential on the free surface, allows the determination of the velocity field on the free surface, which is integrated in time through a two-step Runge-Kutta scheme to update the position of particle of the free surface.
\subsubsection{Pressure distribution}
The pressure distribution along the body contour is obtained through the Bernoulli's equation:
\begin{equation}\label{Eq:8}
p - {p_a} =  - \rho \left({\dot \phi  + \frac{{{{\left| {\nabla \phi } \right|}^2}}}{2}+{\rm g}y} \right),
\end{equation}
and the total hydrodynamic load is obtained by integration of the pressure field along the wetted part of the body
\begin{equation}\label{Eq:9}
\mathbf{F} =  - \int_{{S_B}} {\left( {p - {p_a}} \right){\mathbf{n}}{\rm d}S}.
\end{equation}
$\dot \phi $ has to be provided before the pressure distribution on the wall surface is given. The calculation of $\dot \phi $ is similar to $\phi$. On the free surface, the $\dot \phi $ is known as $\dot \phi  =  - \frac{{{{\left| {\nabla \phi } \right|}^2}}}{2}-{\rm g}y$ according to Eq.$\,$(\ref{Eq:3}). On the wall surface, the normal derivative of $\dot \phi $ is known and calculated as
\begin{equation}\label{Eq:11}
\frac{{\partial \dot \phi }}{{\partial n}} = {\bf a\cdot n} - {w_s}\frac{{\partial {u_n}}}{{\partial s}} - {w_n}\frac{{\partial {u_s}}}{{\partial s}} + {k_s}{\bf W\cdot \mathbf{u}},
\end{equation}
where $\bf a$ is the body acceleration,  $w_s$ and $w_n$ are the tangential and normal projections of ${\bf W}$ on the wall surface, $u_s$ and $u_n$ are the tangential and normal projections of $\mathbf{u}=\nabla \phi$ on the wall surface and $k_s$ denotes the curvature of the wall surface. The normal derivative of $\dot \phi $ on free surface and the $\dot \phi$ on the wedge surface can be solved by a similar boundary integral equation of Eq.$\,$(\ref{Eq:7}). it is worth noting that the gravitational acceleration $\rm g$ is only used for the validations of freefall cases in Figs.$\,$\ref{Fig:CFD_HBF_Wu_comp} and \ref{Fig:freefall_motion_30}, and is never included in other cases.

\subsubsection{Jet model}
\begin{figure}
  \centerline{\includegraphics[width=8cm,height=2.8cm]{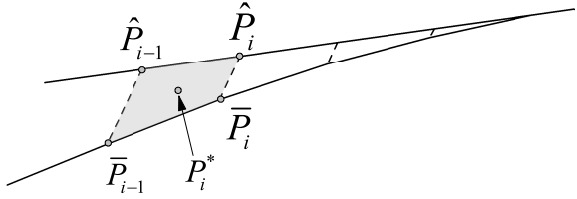}}
  \caption{The description of the jet model of the simplified FEM solver.}
  \label{Fig:chapter8_jet_model}
\end{figure}
In the simplified FEM solver used for the description of the thin jet, a part of the jet region is divided in control volumes in which the vertices corresponding to the panel centroids ($\bar{P}_{i-1}$, $\bar{P}_{i}$, $\hat{P}_{i-1}$, $\hat{P}_{i}$), as shown in \figurename$\,$\ref{Fig:chapter8_jet_model}. In each control volume, the velocity potential is written in the form of a harmonic polynomial expansion, up to second order. Details about the approach in the slamming stage can be found in Battistin and Iafrati \cite{Iafrati2003Hydroelastic} and Del Buono et al. \cite{del2021water}. For the transition stage, the harmonic polynomial expansion, $\phi _i^J$ is reduced to first orders and reads
\begin{equation}\label{Eq:21}
\phi _i^J\left( {x,y} \right) = {\alpha_i} + {\chi_i}\left( {x - x_i^*} \right) + {\mu_i}\left( {y - y_i^*} \right).
\end{equation}
The corresponding normal derivative is
\begin{equation}\label{Eq:22}
\phi _{n,i}^J\left( {x,y} \right) =  {\chi_i}{n_{x,i}}+  {\mu_i}{n_{y,i}}
\end{equation}
where $(x_i^*,\,y_j^*)$ is the centroid of the fluid control volume $P_i^*$, ${n_{x,i}}$ and ${n_{y,i}}$ are the unit vector of the $i^{th}$ panel which are directed along the $x$-axis and $y$-axis, $\alpha_i$, $\chi_i$ and $\mu_i$ are new unknown variables and can be determined by enforcing the free surface condition
\begin{equation}\label{Eq:23}
\phi _i^J\left( {{{\bar P}_{i - 1}}} \right) = \phi \left( {{{\bar P}_{i - 1}}} \right),\quad\phi _i^J\left( {{{\bar P}_i}} \right) = \phi \left( {{{\bar P}_i}} \right),
\end{equation}
and by enforcing the continuity of the $\phi_n$ at adjacent elements
\begin{equation}\label{Eq:20}
\phi _{n,i}^J\left( {{{\bar P}_{i - 1}}} \right) = \phi _{n,i - 1}^J\left( {{{\bar P}_i}} \right).
\end{equation}

\section{Validations of numerical methods}\label{Part:3}

\subsection{Grid independence}\label{Part:3.1}

\begin{figure}
  \centerline{\includegraphics[width=15cm,height=4cm]{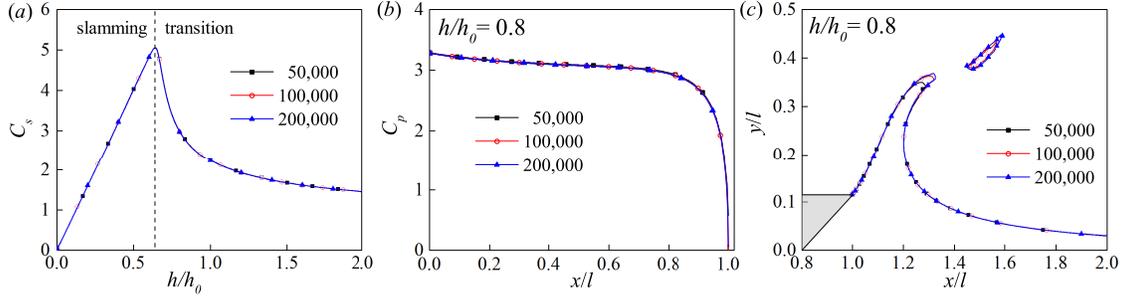}}
  \caption{Grid independence validation of the CFD simulations for the water entry of a wedge with $\beta=30^\circ$ in a constant speed: ($a$) Slamming coefficient $C_s$; ($b$) $C_p$ distribution of $h/h_0=0.8$; Free surface of $h/h_0=0.8$.}
  \label{Fig:CFD_grid_independence}
\end{figure}

Figure$\,$\ref{Fig:CFD_grid_independence} shows the grid independence validation of the CFD simulations for the water entry of a wedge with $\beta=30^\circ$ in a constant speed, where the slamming coefficient  $C_s$ and pressure coefficient $C_p$ are defined as follows
\begin{equation}\label{Eq:Cs_defination}
C_s=\frac{F}{\frac{1}{2}\rho U_w^2 l},
\end{equation}
\begin{equation}\label{Eq:Cp_defination}
C_p=\frac{p-p_a}{\frac{1}{2}\rho U_w^2},
\end{equation}
where $U_w$ is the entering speed of the body. Three grids of cell numbers of 50,000, 100,000 and 200,000 are adopted to simulate the impact flow with adaptive time steps. The courant number is set to be 0.95 during the adaptive time steps. The $C_s$ of the whole time-history between the three grids match well. The $C_p$ distribution and free surface of $h/h_0=0.8$ ($h_0=l\tan\beta$) are also consistent between the three grids except for some discrepancies at the tip of jet. The grid independence of CFD method is successfully validated. Therefore, the CFD grid with cell number of 100,000 is adopted for further simulations by balancing both the accuracy of spatial resolution and the computational cost.

\begin{figure}
  \centerline{\includegraphics[width=13.2cm,height=4.4cm]{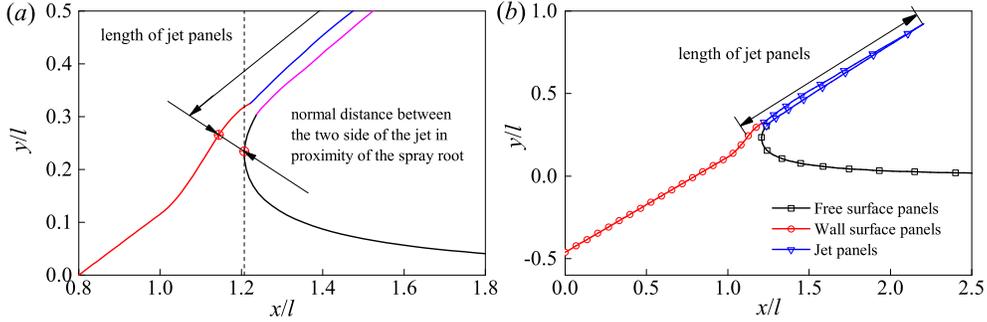}}
  \caption{Description of the BEM panels and FEM panels in the BEM.}
  \label{Fig:description_BEM_FEM}
\end{figure}
\begin{table}
\centering
\begin{tabular}{cccc}
\toprule
Grids&$\kappa$ of BEM panels&$\kappa$ of FEM panels&$r$ \\
\midrule
Coarse&1.04&1.05&1.5\\
Normal &1.03&1.04&2.0\\
Fine&1.02&1.03&4.0\\
\bottomrule
\end{tabular}
\caption{The grid details of BEM panels: $\kappa$ being the growth factor; $r$ being the ratio between the normal distance between the two side of the jet in proximity of the spray root and the minimun panel size.}
\label{Tab:chapter9_BEM_grid_details}
\end{table}

\begin{figure}
  \centerline{\includegraphics[width=15cm,height=5cm]{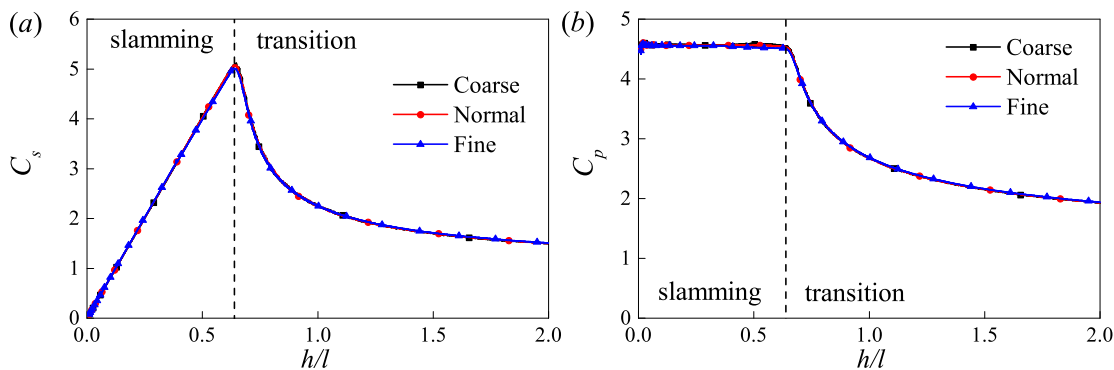}}
  \caption{The grid independence of the HBF simulations for the water entry of a wedge with $\beta=30^\circ$ in a constant speed: ($a$) the slamming coefficient $C_s$; ($b$) the pressure coefficient $C_p$ at the tip of wedge.}
  \label{Fig:BEM_grid_independence}
\end{figure}

Figure$\,$\ref{Fig:description_BEM_FEM} shows the description of the HBF model in which the fluid boundary is divided into three parts: (1) free surface panels; (2) wall surface panels and (3) jet panels. The free surface and wall surface panels are solved by BEM solver while the jet panels are solved by the FEM solver. The panel distribution is determined by three parameters in \tablename$\,$\ref{Tab:chapter9_BEM_grid_details}: growth factors $\kappa$ of BEM panels and FEM panels, as well as the ratio $r$ between the normal distance between the two side of the jet in proximity of the spray root and the minimun panel size. The $C_s$ and $C_p$ at the tip of wedge of the three distributions of HBF for the constant speed case in \figurename$\,$\ref{Fig:BEM_grid_independence} match well with each other. Here, a normal grid is adopted for most of the HBF calculations in present study. 

\subsection{Comparisons between different methods}\label{Part:3.2}

For the slamming stage of constant speed impact in which the self-similar flow is satisfied, the present CFD and HBF methods are compared with the similarity solution of Dobrovol'skaya\cite{dobrovol1969some,wen2020Impact}. For the varying speed impact of both the slamming and transition stages, the present CFD and HBF methods are compared with the predictions of  Bao et al.'s BEM \cite{Bao2017Simulation} for the freefall motion case. In this case, the impact speed is small and the gravity effect will become significant in the transition stage. In the simulations of Bao et al.'s case \cite{Bao2017Simulation}, the gravity of fluid is included. For the other cases, the gravity of fluid is excluded.

\begin{figure}
  \centerline{\includegraphics[width=15cm,height=5cm]{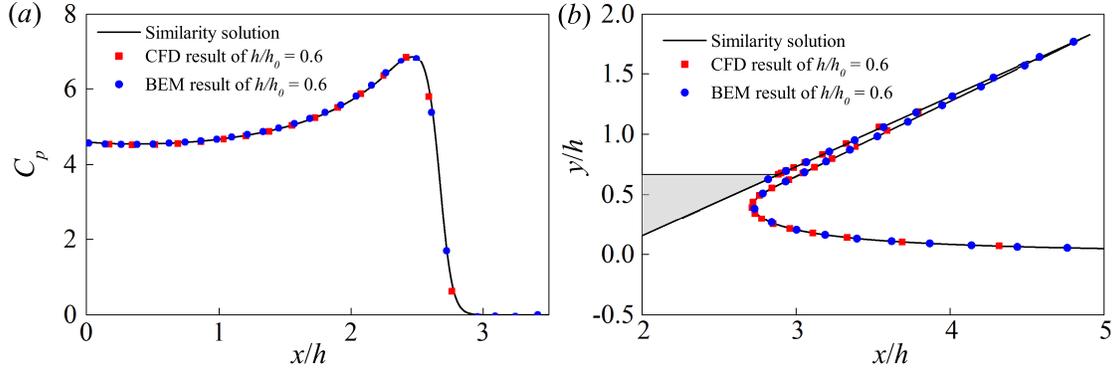}}
  \caption{The comparisons of the pressure coefficient $C_p$ on the wedge surface and the free surface profiles of $h/h_0=0.6$ between the predictions of CFD, HBF and similarity solution \cite{wen2020Impact} during the water entry of a wedge with $\beta=30^\circ$ in a constant speed.}
  \label{Fig:CFD_validations_similarity_solution}
\end{figure}

For the slamming stage with $h/h_0=0.6$ in the case of the water entry of a wedge with $\beta=30^\circ$ in a constant speed, \figurename$\,$\ref{Fig:CFD_validations_similarity_solution} shows the comparisons of the pressure coefficient $C_p$ on the wedge surface and the free surface profiles between the predictions of CFD, HBF and similarity solution \cite{wen2020Impact}. The spray root of the jet remains under the knuckle of the wedge, which means the self-similar flow is still satisfied and thereby the similarity solution can be used for a validation. The good agreement between the present methods and the similarity solution is excellent.

\begin{figure}
  \centerline{\includegraphics[width=15cm,height=4cm]{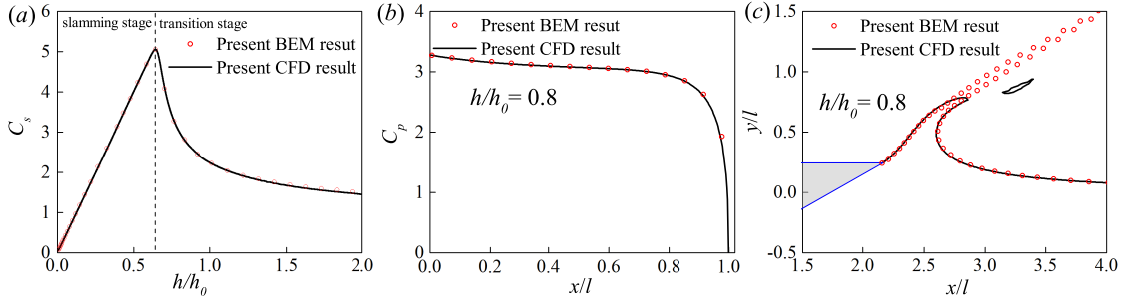}}
 \caption{Comparisons of the force ($a$), pressure distribution ($b$) and free surface ($c$) between the present CFD and HBF for the water entry of a wedge with $\beta=30^\circ$ in a constant speed.}
 \label{Fig:BEM_CFD_compare}
\end{figure}

\begin{figure}
  \centerline{\includegraphics[width=15cm,height=4cm]{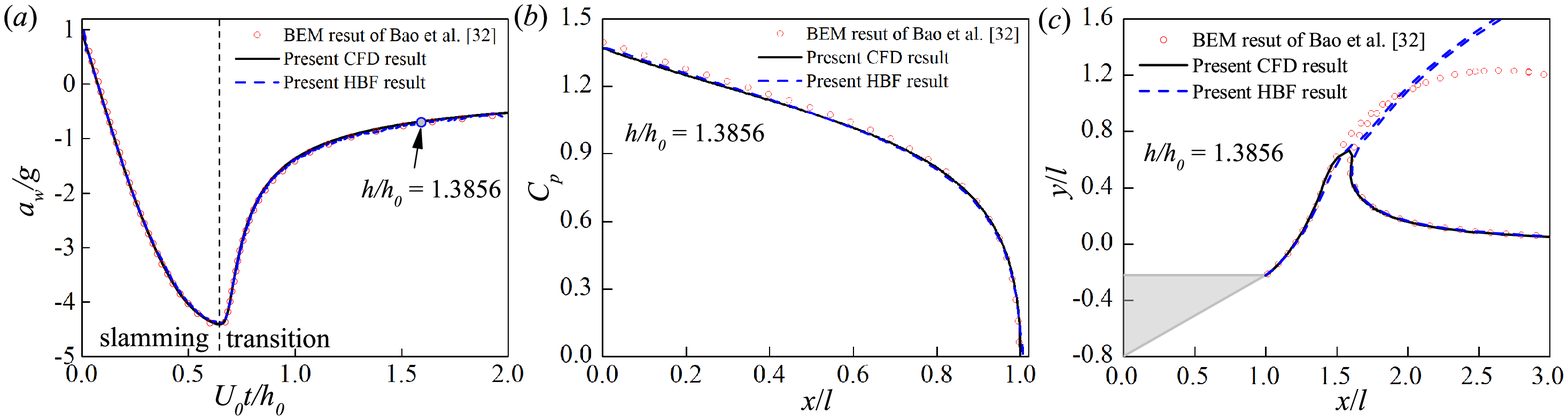}}
 \caption{Comparisons of the acceleration ($a$), pressure distribution ($b$) and free surface ($c$) between the CFD, HBF and BEM of Bao et al. \cite{Bao2017Simulation} for the water entry of wedges of $\beta=30^\circ$ in a freefall motion.}
 \label{Fig:CFD_HBF_Wu_comp}
\end{figure}

For the transition stage, \figurename$\,$\ref{Fig:BEM_CFD_compare} shows the comparisons of the force ($a$), pressure distribution ($b$) and free surface ($c$) between the present CFD and HBF for the water entry of a wedge with $\beta=30^\circ$ in a constant speed. The CFD results are all consistent with those of HBF, except for the jet tip of free surface. Since the jet tip is of little relevance with the pressure and force acting on the wedge surface, the mutual validations between the present CFD and HBF are properly conducted. \figurename$\,$\ref{Fig:CFD_HBF_Wu_comp} shows the comparisons of the acceleration ($a$), pressure distribution ($b$) and free surface ($c$) between the CFD method, HBF and BEM of Bao et al. \cite{Bao2017Simulation} for the water entry of wedges of $\beta=30^\circ$ in a freefall motion. The CFD and HBF results are in good agreement with those of the BEM of Bao et al. \cite{Bao2017Simulation}, while there are still minor discrepancies of the jet tip of free surface and the pressure distribution near the apex of wedge. These differences are acceptable for the validations of CFD and HBF methods.

It can be concluded that the CFD method and HBF method are consistent with each other and can deal with the slamming and transition stages of the constant and varying speed impacts. In this paper, the CFD method is mainly used to produce the numerical results of constant speed impact, while the HBF method is to used produce those of varying speed impact.

\section{Constant speed impact}\label{Part:4}
In this section, the constant speed impacts of wedges with different deadrise angles are studied based on the CFD method. The hydrodynamic forces of the three stages are formulated based on a combination of the results of similarity solution of slamming stage, the CFD results of transition stage and the theoretical results of steady supercavitating flow. The steady supercavitating flow is an approximation of the flow around the wedge at a very large penetration depth and is theoretically formulated in Appendix \ref{App:1} based on the theoretical method of Gurevich \cite{gurevich1965theory}.

\subsection{Pressure distributions of $\beta=30^\circ$ caused by the constant speed impact}\label{Part:4.1}

\begin{figure}
  \centerline{\includegraphics[width=15cm,height=6.8188cm]{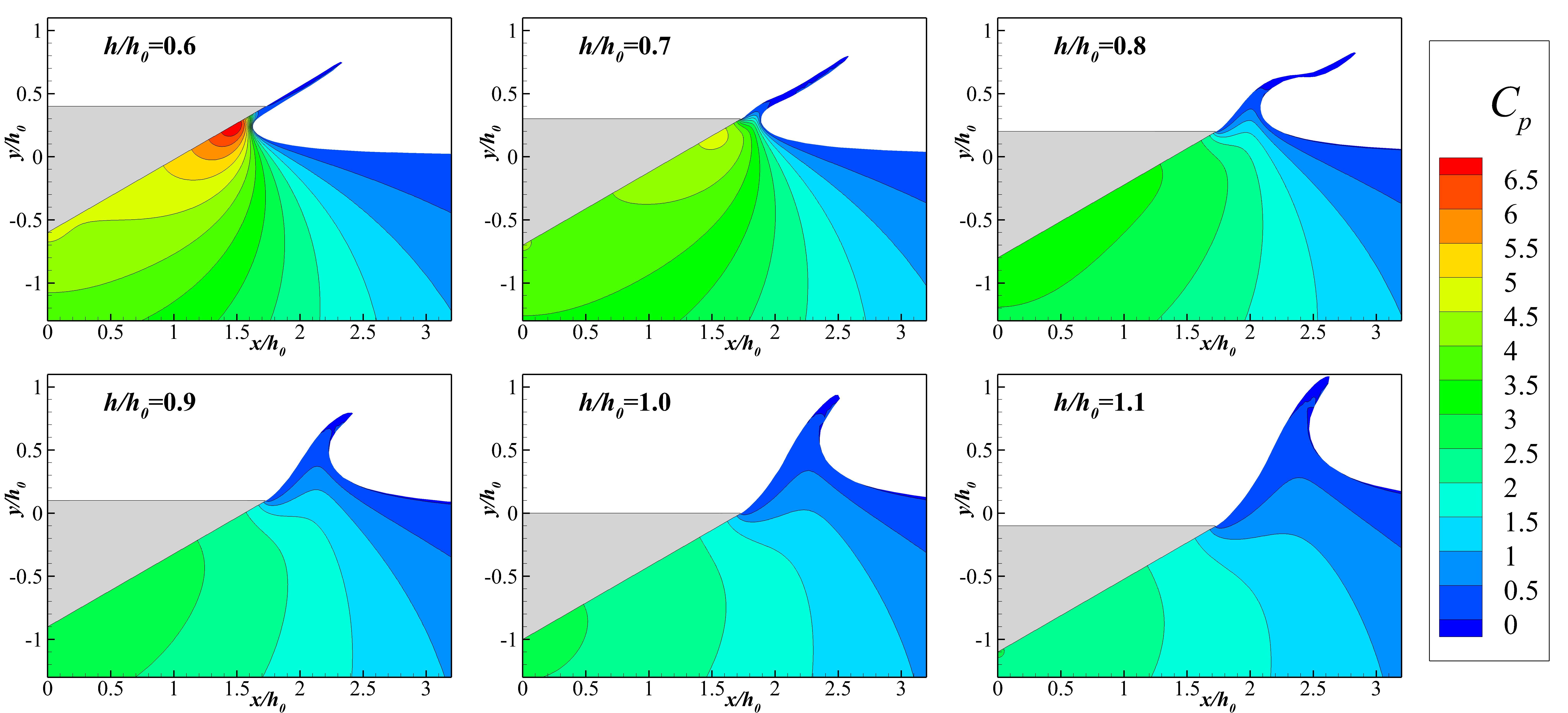}}
  \caption{The pressure distributions and free surface of different penetration depths in the transition stage during the water entry of a wedge with $\beta=30^\circ$ in a constant speed \cite{wen2022Modified}.}
  \label{Fig:transition_stage_Cp_free_surface_contour.jpeg}
\end{figure}

\begin{figure}
  \centerline{\includegraphics[width=10.5cm,height=7cm]{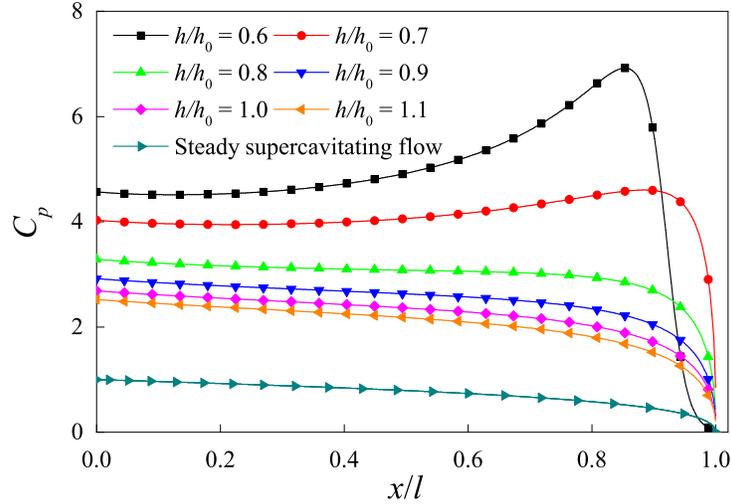}}
  \caption{The pressure distributions on the wedge surface in the transition stage of the water entry of a wedge with $\beta=30^\circ$ in a constant speed. The result of steady supercavitating flow is calculated by the potential theory in Appendix \ref{App:1}.}
  \label{Fig:transition_stage_Cp_on_wedge_surface}
\end{figure}

The pressure distributions and free surface of different penetration depths in the transition stage during the water entry of a wedge with $\beta=30^\circ$ in a constant speed are shown in \figurename$\,$\ref{Fig:transition_stage_Cp_free_surface_contour.jpeg}. It is clearly observed that the high pressure region located at the spray root rapidly vanishes after the spray root leaves the knuckle of the wedge.  In the slamming stage, the high pressure region is formed because the water under the wedge has to accumulate in the spray root and turns into a jet with high speed, as the wetted area of wedge keeps increasing. In the transition stage, the wetted length of the wedge stops increasing and the wall surface of the spray root no longer exists, and thus the high pressure region disappears. \figurename$\,$\ref{Fig:transition_stage_Cp_on_wedge_surface} shows the pressure distributions on the wedge surface. With the increasing penetration depth, the pressure on the whole wedge surface decreases  and finally approaches the distribution of steady supercavitating flow. It is difficult to figure out the way how the pressure drops from the original distribution of slamming stage. In this paper, we focus on the formulation of the force and expect to find a general expression of slamming and transition stages .

\subsection{Hydrodynamic force caused by the constant speed impact}\label{Part:4.2}
\begin{figure}
  \centerline{\includegraphics[width=15cm,height=5cm]{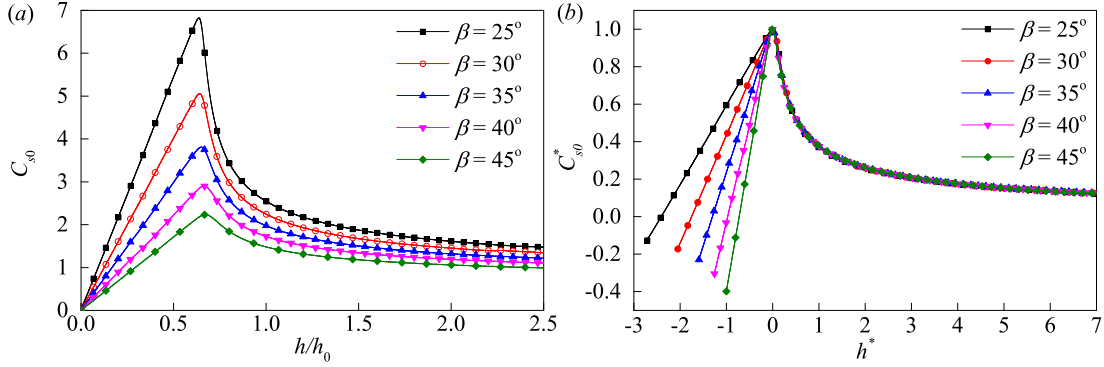}}
  \caption{The slamming coefficient $C_{s0}$ ($a$) and the new variables $C_{s0}^*$ ($b$) of different deadrise angles varying from $25^\circ$ to $45^\circ$ calculated by the CFD method.}
  \label{Fig:slamming_coefficient_original_sorted}
\end{figure}
Figure$\,$\ref{Fig:slamming_coefficient_original_sorted} ($a$) shows the slamming coefficient $C_{s0}$ of different deadrise angles varying from $25^\circ$ to $45^\circ$. The $C_{s0}$ of a single $\beta$ increases linearly to the maximum $C_{\rm max}$ and declines to a steady value $C_{s\infty}$. The $C_{s\infty}$ of different deadrise angles are shown in \figurename$\,$\ref{Fig:Cs0_infinite_stage} in Appendix \ref{App:1}, which is calculated by a potential theory \cite{gurevich1965theory}. In order to formulate the slamming coefficient $C_{s0}$ in the transition stage, new variables are adopted:
\begin{equation}\label{Eq:SC1}
h^*=\frac{h-h_2}{h_2}\cot^{\lambda}\beta,
\end{equation}
\begin{equation}\label{Eq:SC2}
C_{s0}^*=\frac{C_{s0}-C_{s\infty}}{C_{s\rm max}-C_{s\infty}},
\end{equation}
where $h_2$ is the penetration depth corresponding to the maximum slamming coefficient $C_{s\rm max}$. The new parameter $\lambda$ is determined by a gradient algorithm enforcing on the following function
\begin{equation}\label{Eq:SC3}
{\eta}(\lambda) = \frac{1}{N}\sum\limits_{i = 1}^N {\sigma _i^2(\lambda)},
\end{equation}
where $\sigma_i (\lambda)$ is the standard deviation of $C_{s0}^*$ between different deadrise angles for the $i^{th}$ node in the range of $h^*\in[0,\,\,7]$ and thereby becomes a function of $\lambda$. An uniform distribution with $N$=7001 nodes in the range of $h^*\in[0,\,\,7]$ is adopted, and the initial values are chosen as $\lambda=1.4$ and $1.35$. After 35 steps, the gradient algorithm is convergent with $\lambda=1.3075$ ($\Delta \lambda<$1e-6). The $C_{s0}^*$ of different deadrise angles with $\lambda=1.3075$ are also shown in \figurename$\,$\ref{Fig:slamming_coefficient_original_sorted} ($b$), and it can be seen that the $C_{s0}^*$ of different deadrise angles in the transition stage coincide well  with each other. 

\begin{figure}
  \centerline{\includegraphics[width=10cm,height=5cm]{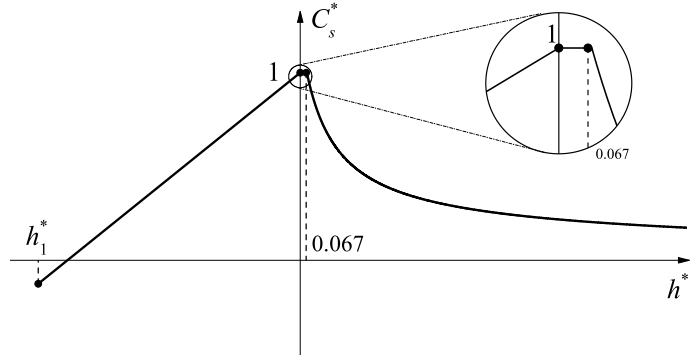}}
  \caption{Description of the formulation for the whole process of water entry.}
  \label{Fig:Csstar_hstar_description}
\end{figure}
\begin{figure}
  \centerline{\includegraphics[width=15cm,height=5cm]{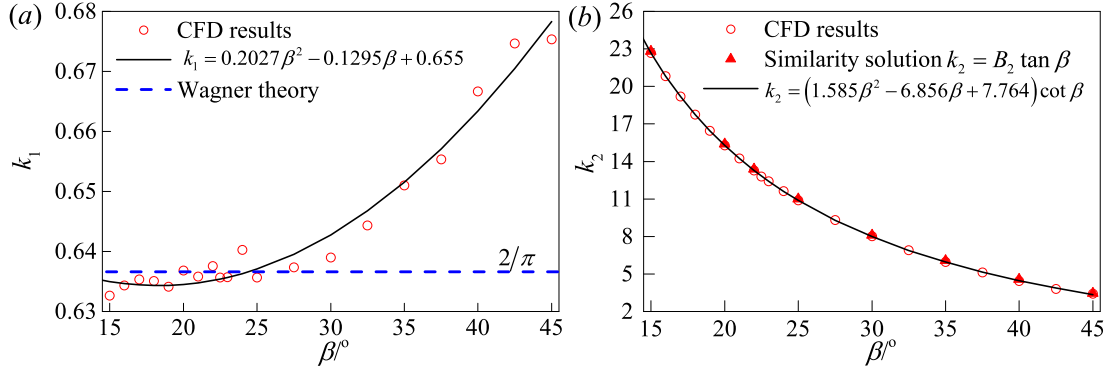}}
  \caption{The $k_1$ and $k_2$  of various deadrise angles from 15$^\circ$ to 45$^\circ$, which are calculated by the present CFD method.}
  \label{Fig:maximum_Cs_h0_BEM_CFD}
\end{figure}

Owing to the linear increasing slamming coefficient in the slamming stage and the coinciding results in the transition stage, the hydrodynamic force of both the slamming and transition stages of water entry can be formulated as shown in \figurename$\,$\ref{Fig:Csstar_hstar_description}, where  the slamming stage is in the range of  $[h_1^*,\,\,0]$ and the transition stage is in that of $[0,+\infty]$. The coinciding $C_{s0}^*$ of \figurename$\,$\ref{Fig:slamming_coefficient_original_sorted} ($b$) in the transition stage is formulated by a rational function in $h^*\in[0,\,\,7]$ based on the mean results of the $C_{s0}^*$ of different deadrise angles
\begin{equation}\label{Eq:SC6}
C_{s0}^* = \frac{{1.539{h^*} + 2.618}}{{{h^{*2}} + 8.081{h^*} + 2.169}}.
\end{equation}
Due to the fitting errors, $C_{s0}^*$ in Eq.$\,$(\ref{Eq:SC6}) is larger than 1 when $h^*\in[0,\,\,0.067]$. In order to deal with the continuity between Eq.$\,$(\ref{Eq:SC6}) and the linear increasing $C_{s0}^*$ of slamming stage, $C_{s0}^*$ of  is approximately as 1 in $h^*\in[0,\,\,0.067]$. The linear increasing $C_{s0}^*$ appears the following form
\begin{equation}\label{Eq:SC5}
C_{s0}^* ={1 + \frac{{\partial C_{s0}^*}}{{\partial {h^*}}}{h^*}}.
\end{equation}
By taking the derivative of $C_{s0}^*$ with respect to $h^*$, the slope of the linear expression of Eq.$\,$(\ref{Eq:SC5}) in $[h_1^*,\,\,0]$ appears to be
\begin{equation}\label{Eq:SC4}
\frac{{{\rm d} C_{s0}^*}}{{{\rm d} {h^*}}} = \frac{{{k_1}{k_2}{{\tan }^{\lambda}}\beta }}{{{C_{s\max }} - {C_{s\infty}}}},
\end{equation}
where $k_1=h_2/h_0$, $k_2=\frac{{\rm d}C_{s0}}{{\rm d} (h/h_0)}$.  $C_{s\max }=k_1k_2$ according to the definations of $k_1$ and $k_2$.
\figurename$\,$\ref{Fig:maximum_Cs_h0_BEM_CFD} shows the $k_1$ and $k_2$ of various deadrise angles from 15$^\circ$ to 45$^\circ$, which are calculated by the present CFD method. In the Wagner theory, $k_1={2}/{\pi}$. $k_1$ from CFD results match $2/\pi$ when $\beta\leq30^\circ$. Thus, $k_1=2/\pi$ can be adopted for small deadrise angle. $k_2$ appears to be equal to $B_2\tan\beta$ of the similarity solution \cite{dobrovol1969some, wang2017improved,wen2020Numerical}. The CFD results of $k_2$ are in good agreement with the similarity solution. In this paper, the results of $k_1$ and $k_2$ for an arbitrarily deadrise angle within the range of $[15^\circ,\,\,45^\circ]$ are given by a least square fitting method (LSF) of a quadratic function ${k_1} = 0.2027{\beta ^2} - 0.1295\beta  + 0.655$ and a quadratic function multiplying $\cot\beta$, e.g., ${k_2} = \left( {1.585{\beta ^2} - 6.856\beta  + 7.764} \right)\cot \beta $ from the CFD results, and the maximum fitting errors are $0.62\%$ and $1.91\%$ respectively. Therefore
\begin{equation}\label{Eq:total_constant}
C_{s0}^* = \left\{ {\begin{array}{*{20}{c}}
{1 + \frac{{\partial C_{s0}^*}}{{\partial {h^*}}}{h^*},}&{{h^*} \le 0}\\
{\begin{array}{*{20}{c}}
{1,}\\
{\frac{{1.539{h^*} + 2.618}}{{{h^{*2}} + 8.081{h^*} + 2.169}},}
\end{array}}&{\begin{array}{*{20}{c}}
{0 < {h^*} \le 0.067}\\
{{h^*} > 0.067}
\end{array}}
\end{array}} \right.
\end{equation}

By combining the Eqs.$\,$(\ref{Eq:SC1}) -  (\ref{Eq:SC2}) and (\ref{Eq:total_constant}), it is possible to find the $C_{s0}$ value and then the hydrodynamic force acting on the wedge surface (half model) for the cases in a constant speed is given
\begin{equation}\label{Eq:SC8}
F = \frac{1}{2}\rho U_w^2l{C_{s0}}.
\end{equation}

\begin{comment}
\begin{table}
\centering
\begin{tabular}{cccc}
\toprule
Variable&Fitting function &Error\\
\midrule
$k_1$&$0.2027{\beta ^2} - 0.1295\beta  + 0.655$&$0.62\%$\\
${k_2}$&$\left( {1.585{\beta ^2} - 6.856\beta  + 7.764} \right)\cot \beta $&$1.91\%$\\
%$A_2$&$\left( {0.844{\beta ^2} - 2.937\beta  + 3.838} \right)\cot^2 \beta $&Similarity solution&$2.14\%$\\
\bottomrule
\end{tabular}
\caption{Fitting functions of $k_1$ and $k_2$ and the corresponding fitting errors.}
\label{Tab:fitting functions}
\end{table}
\end{comment}

\subsection{Comparisons with CFD results}\label{Part:4.3}

In Sec.$\,$\ref{Part:4.2}, the formulation of slamming stage in the slamming and transition stages is set up by Eqs.$\,$(\ref{Eq:SC1})-(\ref{Eq:SC2}) and (\ref{Eq:total_constant}), and the hydrodynamic force is finally calculated by Eq.$\,$(\ref{Eq:SC8}). The formulation is based on the CFD results in a range of deadrise angles from $15^\circ$ to $45^\circ$. In order to explore its application range of deadrise angles, this section compares the predictions of  Eq.$\,$(\ref{Eq:SC8}) and the CFD results in the range of $\beta\in [5^\circ,70^\circ]$.

\begin{figure}
  \centerline{\includegraphics[width=15cm,height=4cm]{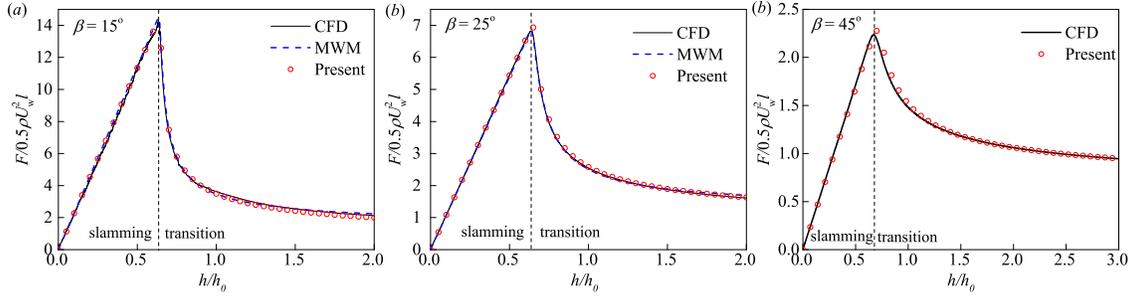}}
  \caption{The comparisons between the predictions of Eq.$\,$(\ref{Eq:SC8}), the modified Wagner's model (MWM) \cite{wen2022Modified} and  the CFD method  for the water entry of linear wedges with deadrise angles of $\beta=15^\circ$, $25^\circ$ and $45^\circ$.}
  \label{Fig:Cs_all_angles_predictions}
\end{figure}

For the comparisons in $\beta\in [15^\circ,\,\,45^\circ]$, \figurename$\,$\ref{Fig:Cs_all_angles_predictions} shows the comparisons between the predictions of Eq.$\,$(\ref{Eq:SC8}), the modified Wagner's model (MWM) \cite{wen2022Modified} and  the CFD method  for the cases of $\beta=15^\circ$, $25^\circ$ and $45^\circ$. The agreement between the present model and the CFD results is good except that a tiny discrepancy occurs at the maximum $C_s$ for the case of $\beta=45^\circ$. The WMW can only work in the range of $\beta\in[15^\circ, \,\,35^\circ]$.  For the cases of $\beta=15^\circ$ and $25^\circ$, the accuracy of the present model and MWM is close. 

\begin{figure}
  \centerline{\includegraphics[width=15cm,height=5cm]{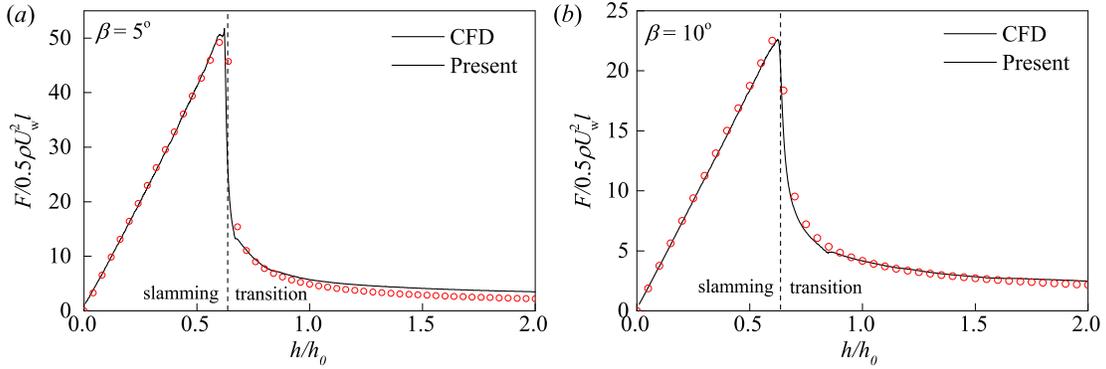}}
  \caption{The comparisons between the predictions of Eq.$\,$(\ref{Eq:SC8}) and  the CFD method  for water entry of linear wedges with small deadrise angles of $\beta=5^\circ$ and $10^\circ$.}
  \label{Fig:Low_angles_predictions}
\end{figure}

\begin{figure}
  \centerline{\includegraphics[width=15cm,height=4cm]{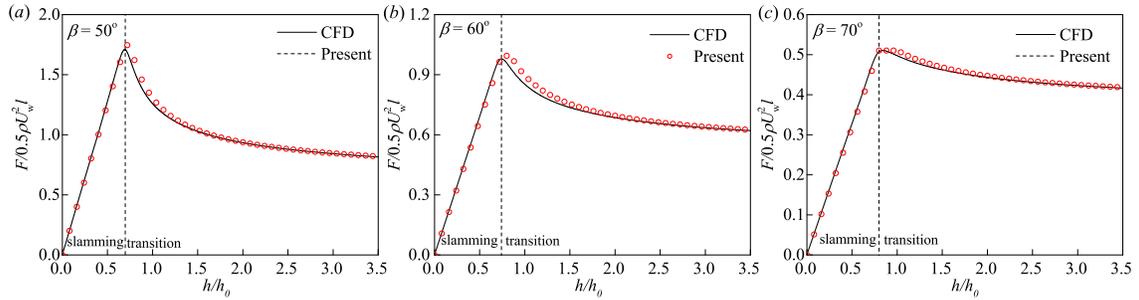}}
  \caption{The comparisons between the predictions of Eq.$\,$(\ref{Eq:SC8}) and  the CFD method  for water entry of linear wedges with high deadrise angles of  $\beta=50^\circ$, $60^\circ$ and $70^\circ$.}
  \label{Fig:High_angles_predictions}
\end{figure}

The comparisons in the ranges of small and large deadrise angles are also provided in Figs.$\,$\ref{Fig:Low_angles_predictions} and $\,$\ref{Fig:High_angles_predictions}. The agreement between the predictions of Eq.$\,$(\ref{Eq:SC8}) and the CFD results is generally good, though with some discrepancies. Eq.$\,$(\ref{Eq:SC8}) slightly underestimates the force at a large penetration depth for the case of $\beta=5^\circ$ and overestimates the forces for the cases of large deadrise angles at an early period of the transition stage. But the agreement becomes better as the penetration depth increases. Therefore, it can be concluded that the present formula can provide accurate predictions for the slamming coefficients of various deadrise angles from $5^\circ$ to $70^\circ$. 

Since there exists entrainment of air when the deadrise angle is smaller than $5^\circ$ \cite{chuang1966experiments,chuang1969investigation}, the predictions of force without involvement of the effect of air cushion may be inaccurate. Thus, this paper does not consider the validations of a smaller deadrise angle, and $5^\circ\le \beta\le 70^\circ$ will be a proper range of deadrise angle for most of the engineering applications.


\section{Varying speed impact}\label{Part:5}

In this section, the varying speed impact of a linear wedge is numerically studied by the HBF method described in Sec.$\,$\ref{Part:2.2}. The varying speed cases are shown in \tablename$\,$\ref{Tab:chapter9_presecribed_speed}. The wedges have the same half-width of 1m. All the cases start the varying speed motions at the start time in the slamming stage, which makes sure that the impacts in the transition stage are all in varying speed motions. Cases 1 and 4 have constant decelerations and the other have linear decelerations. In the following simulations of HBF, the growth factor of BEM and FEM panels are 1.03 and 1.04 respectively and the ratio between the normal distance between the two side of the jet in proximity of the spray root (to be honest where the jet model start) and the minimun panel size is 2. The pressure distributions and the force acting on the wedge body are discussed by comparing the results of the constant and varying speed impacts. The formulation of the acceleration effect in the transition stage is proposed and the new expression will be validated by numerical and experiment results and compared with other theories.
\begin{comment}
\begin{table}
\centering
\begin{tabular}{ccccccc}
\toprule
Cases&Case 1&Case 2&Case 3&Case 4&Case5&Case 6\\
\midrule
$\beta\,\rm (^\circ)$&30&30&30&14&20&45\\
$t_0$ (s)&0.25&0.25&0.25&0.06&0.15&0.63\\
$U_0\,\rm (m/s)$&1&1&1&2&1&1\\
$a_w\,\rm (m/s^2)$&-1&-2$t$+0.2&$2t-1$&$-5$&$-2t+0.3$&$-0.5t+0.315$\\
\bottomrule
\end{tabular}
\caption{Details of the cases varying speed impacts in forced motions. The cases remains the impacts in a constant speed of $U_0$ before the start time and turn into those of a varying speed with the above accelerations after that. Cases 1 and 4 have constant decelerations and the other have linear decelerations.}
\label{Tab:chapter9_presecribed_speed}
\end{table}
\end{comment}
%\begin{comment}
\begin{table}
\centering
\begin{tabular}{ccccc}
\toprule
&$\beta\,\rm (^\circ)$&  $U_0\,\rm (m/s)$ &start time (s)& $a_w\,\rm (m/s^2)$\\
\midrule
Case 1&30&1&0.25&$-1$\\
Case 2&30&1&0.25&$$-2t+0.2$$\\
Case 3&30&1&0.25&$2t-2$\\
Case 4&15&2&0.06&$-5$\\
Case 5&20&1&0.15&$-2t+0.3$\\
Case 6&45&1&0.63&$-0.5t+0.315$\\
\bottomrule
\end{tabular}
\caption{Details of the cases of varying speed impacts in forced motions. The cases remains the impacts in a constant speed of $U_0$ before the start time and turn into those of a varying speed with the above accelerations after that. Cases 1 and 4 have constant decelerations and the other have linear decelerations.}
\label{Tab:chapter9_presecribed_speed}
\end{table}
%\end{comment}

\subsection{Pressure distributions of $\beta=30^\circ$ caused by the acceleration effect}\label{Part:5.1}
\begin{figure}
  \centerline{\includegraphics[width=15cm,height=5cm]{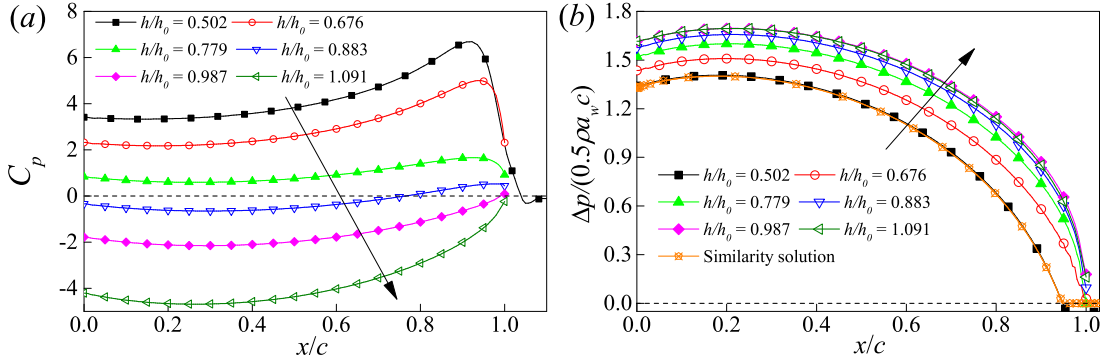}}
  \caption{The $C_p$ distributions on the wedge surface in the transition stage of Case 1 (see \tablename$\,$\ref{Tab:chapter9_presecribed_speed}) ($a$) and the distributions of $\Delta p/(0.5\rho a_wc)$, where $\Delta p$ is the difference of the pressure  between the varying and constant speed impacts in a same penetration depth and with a same instantaneous speed ($b$). The result of similarity solution $k_1\tan\beta(-2\Phi)$ is also given for comparisons\cite{wen2020Impact}. The result of $h/h_0=0.502$ remains in the slamming stage.}
  \label{Fig:Varying_speed_pressure_distribution}
\end{figure}
Figure$\,$\ref{Fig:Varying_speed_pressure_distribution} ($a$) shows the $C_p$ distributions on the wedge surface in the transition stage of Case 1 (see \tablename$\,$\ref{Tab:chapter9_presecribed_speed}). For the case of $\beta=30^\circ$, the transition stage starts at $h/h_0=0.639$ and thus the result of $h/h_0=0.502$ is still in the slamming stage while the other are in the transition stage. As can be seen in \figurename$\,$\ref{Fig:Varying_speed_pressure_distribution} ($a$), the $C_p$ distributions along the wedge surface decline with the increasing penetration depth. The reasons of the pressure drop includes two different aspects: the vanishing of high pressure region at spray root (see Sec.$\,$\ref{Part:3.1}) and the deceleration of wedge. In order to distinguish these two kinds of influence, the pressure is split into two parts: (1) the pressure only related to the constant speed impact $p_0=0.5\rho U_w^2C_{p0}$ (see \figurename$\,$\ref{Fig:transition_stage_Cp_on_wedge_surface}); (2) the pressure changing $\Delta p=p-p_0$ caused by the acceleration of wedge, where $p_0$ and $p$ are the pressure of the constant and varying speed impacts respectively in a same penetration depth and with a same instantaneous speed. \figurename$\,$\ref{Fig:Varying_speed_pressure_distribution} ($b$) shows the distributions of $\Delta p/\left(0.5\rho a_wc\right)$, where the effective wetted length (half length) is $c = \min \left\{ {h\cot \beta /{k_1},\,\,l} \right\}$. The $\Delta p/\left(0.5\rho a_wc\right)$ of different penetration depths show similar distributions and have a small increase along the whole wedge surface with the increasing penetration depth. The result of similarity solution $k_1\tan\beta(-2\Phi)$ \cite{wen2020Impact} in the slamming stage is well consistent with the result $h/h_0=0.502$ in the slamming stage. It can be concluded that the distribution of $\Delta p/\left(0.5\rho a_wc\right)$ remains as $k_1\tan\beta(-2\Phi)$ in the slamming stage and then has a small increase in the transition stage. The results become steady as the penetration depth continues to increase. The small increase  pressure will also result in the small increase of slamming coefficient, which will be further discussed in Sec. \ref{Part:5.2}.

\begin{figure}
  \centerline{\includegraphics[width=12cm,height=8cm]{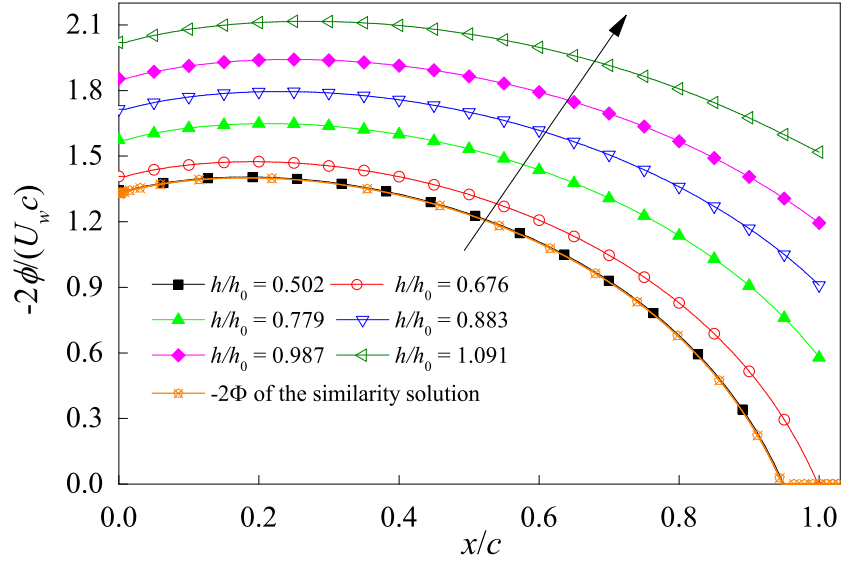}}
  \caption{The dimensionless velocity potential $-2\phi/(U_wc)$ on the wedge surface of Case 1 in the transition stage. The result of $h/h_0=0.502$ remains in the slamming stage.}
  \label{Fig:velocity_potential_distributions_varying_speed}
\end{figure}

From the studies of slamming stage \cite{wen2020Impact,wen2020Numerical}, the pressure changing caused by the acceleration of wedge can be quantified by the dimensionless velocity potential $-2\Phi$. \figurename$\,$\ref{Fig:velocity_potential_distributions_varying_speed} shows the dimensionless velocity potential $-2\phi/\left(U_wc\right)$ (a similar dimensionless variable like $-2\Phi$) on the wedge surface of Case 1 in the transition stage. The $-2\phi/(U_wc)$ distribution of $h/h_0=0.502$ is identical to $-2\Phi$ of the similarity solution \cite{wen2020Numerical} since it still remains in the slamming stage. The other distributions show large differences with $-2\Phi$ of the similarity solution and the $\Delta p/\left(0.5\rho a_wc\right)$ distributions in \figurename$\,$\ref{Fig:Varying_speed_pressure_distribution} ($b$). It can be concluded that the pressure changing caused by the acceleration of wedge can not be quantified by the velocity potential in the transition stage as it does in the slamming stage. Although the $\Delta p/\left(0.5\rho a_wc\right)$ distributions have a similar distribution and small increase as the penetration depth increases,  there is still some challenging problems to formulate the pressure changing caused by the acceleration effect. In this paper, we focus on the formulation of the hydrodynamic force caused by the acceleration effect and hope to find an effective method to formulate the hydrodynamic force caused by the acceleration effect.

\subsection{Hydrodynamic force caused by the acceleration effect}\label{Part:5.2}
\begin{figure}
  \centerline{\includegraphics[width=12cm,height=7.2cm]{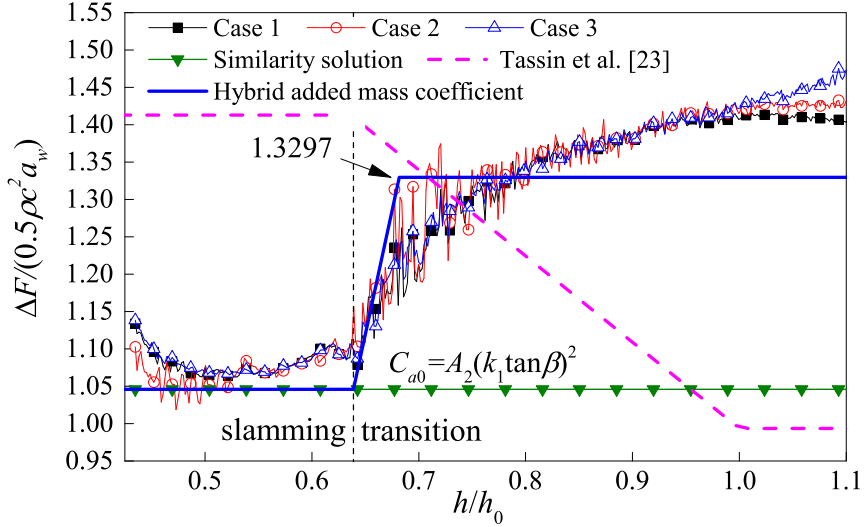}}
  \caption{The time histories of $\Delta F/(0.5\rho c^2a_w)$ of different prescribed speeds (Case 1, Case 2 and Case 3 in \tablename$\,$\ref{Tab:chapter9_presecribed_speed}), where $\Delta F$ is difference of force between the varying and constant speed impacts in a same penetration depth and with a same instantaneous speeds. The results of $C_{a0}=A_2(k_1\tan\beta)^2$ from the similarity solution  \cite{wen2020Impact} and Eq.$\,$(\ref{Eq:Ca_Tassin_trans}) from Tassin et al. \cite{Tassin2014On} are also given for comparisons. }
  \label{Fig:difference_force_varying_constant}
\end{figure}
Similar to the pressure distribution on the wedge surface, the force can also be divided into two part:
(1) the force caused by the constant speed impact $F_0=0.5\rho U_w^2l C_{s0}$, where $C_{s0}$ can be given by Eqs.$\,$(\ref{Eq:SC1}) -  (\ref{Eq:SC2}) and (\ref{Eq:total_constant}); (2) the force caused by the acceleration effect $\Delta F=F-F_0$, where $F_0$ and $F$ are the forces of constant and varying speed impacts respectively in a same penetration depth and with a same instantaneous speed. \figurename$\,$\ref{Fig:difference_force_varying_constant} shows the $\Delta F/(0.5\rho c^2 a_w)$ of different prescribed speeds (Cases 1,2 and 3 in \tablename$\,$\ref{Tab:chapter9_presecribed_speed}). All of the cases are decelerating with different decelerations. Case 1 has a constant deceleration, Case 2 has an increasing deceleration and Case 3 has a decreasing deceleration. The $\Delta F/(0.5\rho c^2 a_w)$ show good consistence between different cases in the transition stage of $h/h_0\in[0.639\,\,1.0]$, though with some numerical fluctuations. Therefore, $C_a=\Delta F/(0.5\rho c^2 a_w)$ is defined as an added mass coefficient, and independent of $a_w$ in $h/h_0\in[0.639\,\,1.0]$. It can be concluded that the effects of acceleration on the hydrodynamic force have a fixed pattern in the transition stage as it does in the slamming stage \cite{wen2020Numerical}. In the slamming stage, the acceleration effect on the hydrodynamic force is given as $\Delta F=\rho A_2a_wh^2$ from the similarity solution \cite{wen2020Impact}, indicating an added mass coefficient $C_{a0}=A_2(k_1\tan\beta)^2$. The $C_a$ in the slamming stage can be given as $C_{a0}=1.0459$, and in the transition stage, $C_a$ has an averaged result of 1.3297 in $h/h_0\in[0.639\,\,1.0]$, which is $\xi=27.13\%$ larger than the result of similarity solution $C_{a0}$. In this paper, a hybrid added mass coefficient is adopted to formulate the acceleration effect corresponding to Eq.$\,$(\ref{Eq:total_constant})
\begin{equation}\label{Eq:SV1}
{C_a} = \left\{ {\begin{array}{*{20}{c}}
{{C_{a0}},}&{{h^*} \le 0}\\
{\begin{array}{*{20}{c}}
{(1 + \xi {h^*}/0.067){C_{a0}},}\\
{(1 + \xi ){C_{a0}}}
\end{array}}&{\begin{array}{*{20}{c}}
{0 < {h^*} \le 0.067}\\
{{h^*} > 0.067}
\end{array}}
\end{array}} \right.
\end{equation}
The results of $h^*\in [0,\,0.067]$ are given by a linear distribution from $C_{a0}$ to  $(1+\xi)C_{a0}$. For the case of $\beta=30^\circ$, the approximation of $C_a$ in the transition stage is summarized from the HBF results of $h/h_0\in[0.639,\,1.0]$, corresponding to $h^*\in[0,\,1.1586]$. However, further validations of  Eq.$\,$(\ref{Eq:SV1}) by the numerical results in Sec.$\,$\ref{Part:5.3} indicates that Eq.$\,$(\ref{Eq:SV1}) can work in a larger range of $h^*$ and deadrise angles. Therefore, for the water entry of linear wedges with an acceleration, the hydrodynamic force can be predicted by Eq.$\,$(\ref{Eq:SV1})   together with Eqs.$\,$(\ref{Eq:SC1}) -  (\ref{Eq:SC2}) and (\ref{Eq:total_constant}), and the final expression of the force (half model) has the following form:
\begin{equation}\label{Eq:SV2}
F = \frac{1}{2}\rho U_w^2l{C_{s0}} + \frac{1}{2}\rho {c^2}{a_w}{C_a},
\end{equation}
where $c = \min \left\{ {h\cot \beta /{k_1},\,\,l} \right\}$.

 In the linear FBC of Tassin et al. \cite{Tassin2014On} based on MLM, the acceleration effect was addressed by the following form of $C_a$ of  in the slamming and transition stages 
\begin{equation}\label{Eq:Ca_Tassin_trans}
{C_a} = \left\{ {\begin{array}{*{20}{c}}
{\frac{\pi }{2} + \left( {1 - \frac{4}{\pi }} \right)\tan \beta ,}&{\frac{h}{{{h_0}}} \le \frac{2}{\pi }}\\
{\begin{array}{*{20}{c}}
{\frac{\pi }{2} + \left( {1 - \frac{{2h}}{{{h_0}}}} \right)\tan \beta ,}\\
{\frac{\pi }{2} - \tan \beta. }
\end{array}}&{\begin{array}{*{20}{c}}
{\frac{2}{\pi } < \frac{h}{{{h_0}}} \le 1}\\
{\frac{h}{{{h_0}}} > 1}
\end{array}}
\end{array}} \right.
\end{equation}
As can be seen in \figurename$\,$\ref{Fig:difference_force_varying_constant}, the present model has a different $C_a$ from that of Tassin et al. \cite{Tassin2014On}.

\subsection{Validations by numerical and experiment results}\label{Part:5.3}

\begin{figure}
  \centerline{\includegraphics[width=15cm,height=4cm]{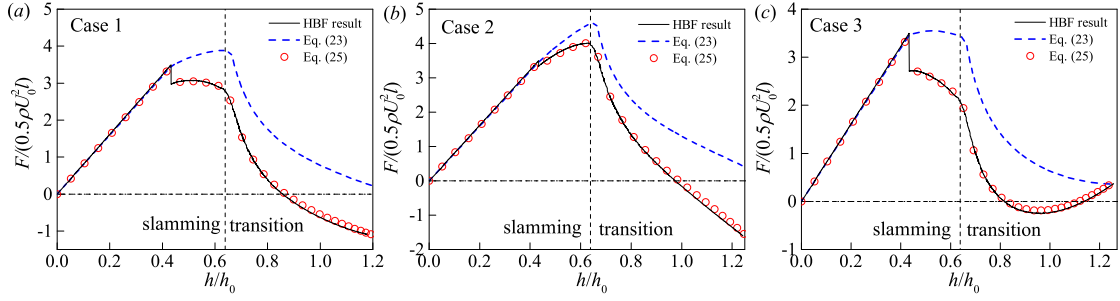}}
  \caption{The comparisons of the forces acting on the wedge surface between the present predictions and the HBF results for Case 1, 2 and 3.}
  \label{Fig:predictions_30deg_CasesABC}
\end{figure}

\begin{figure}
  \centerline{\includegraphics[width=15cm,height=4cm]{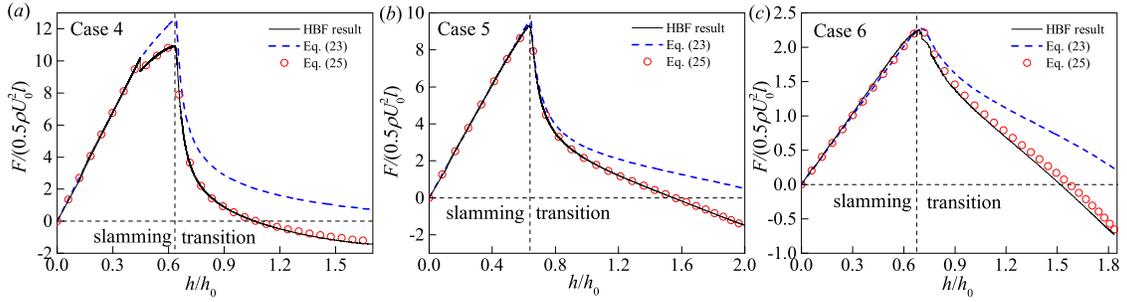}}
  \caption{The comparisons of the forces acting on the wedge surface between the present predictions and the HBF results for Case 4, 5 and 6.}
  \label{Fig:predictions_14_20_45_CasesDEF}
\end{figure}

Figure$\,$\ref{Fig:predictions_30deg_CasesABC} shows the comparisons of the forces acting on the wedge surface between the present predictions and the HBF results for Case 1, 2 and 3. The results of Eq.$\,$(\ref{Eq:SC8}) only include the effect of constant speed impact while those of Eqs.$\,$(\ref{Eq:SV2}) include the effect of constant speed impact and the acceleration effect. The predictions of Eqs.$\,$(\ref{Eq:SV2}) are in good agreement with the HBF results, while those of Eqs.$\,$(\ref{Eq:SC8}) show much discrepancies with the HBF results. \figurename$\,$\ref{Fig:predictions_14_20_45_CasesDEF} shows comparisons for Case 4, 5 and 6 of different deadrise angles and different ways of decelerations. The same $C_a$ correction of $\xi=27.13\%$ is adopted for the cases of different deadrise angles, and the good agreement is also obtained. It indicates that the  $C_a$ correction of $\xi=27.13\%$ also works for other deadrise angles.


\begin{figure}
  \centerline{\includegraphics[width=15cm,height=5cm]{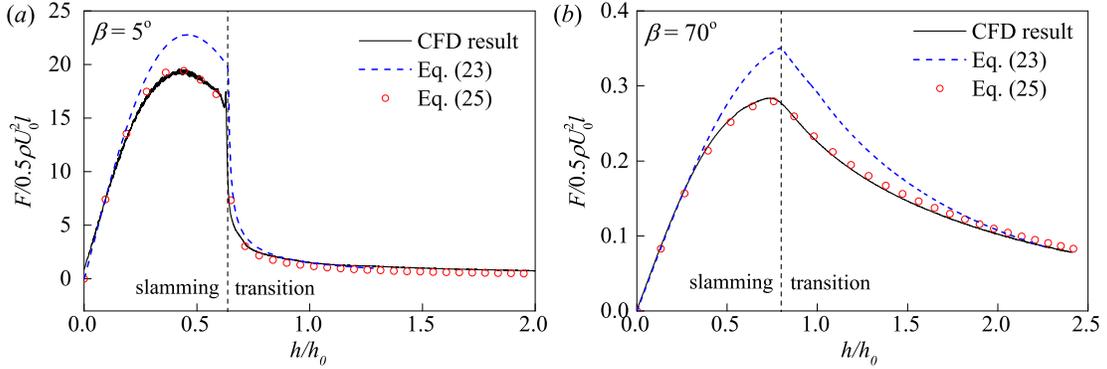}}
  \caption{The comparisons of the forces acting on the wedge surface between the present predictions and the CFD results for the freefall cases of ($a$) $\beta=5^\circ$ and ($b$) $\beta=70^\circ$ with masses of $m=1.5\rho l^2$.}
  \label{Fig:Low_high_angles_acc}
\end{figure}

To explore the application ranges of deadrise angles and of different ways of wedge motions for the present model, \figurename$\,$\ref{Fig:Low_high_angles_acc} shows the comparisons of the forces acting on the wedge surface between the present predictions and the CFD results for the freefall cases of ($a$) $\beta=5^\circ$ and $\beta=70^\circ$. The wedges are in freefall motions with masses of $m=1.5\rho l^2$. For the small and high deadrise angles, the present model can provide good predictions for the hydrodynamic forces, even for the freefall motions where the acceleration of wedge is unknown before the predictions.

\begin{figure}
  \centerline{\includegraphics[width=9cm,height=6cm]{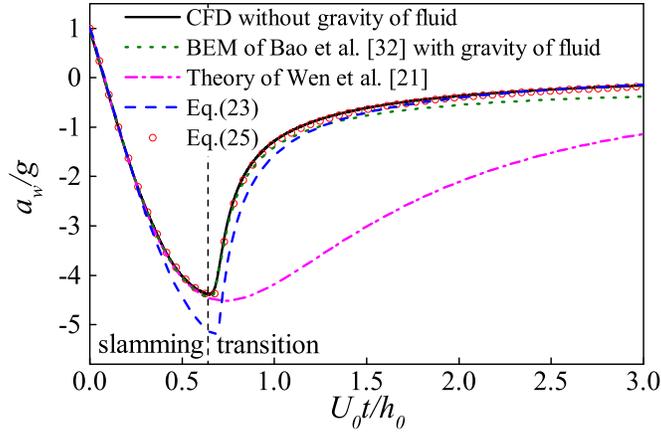}}
  \caption{The comparisons of the predictions of accelerations of wedge with $\beta=30^\circ$ between the CFD without gravity of fluid, BEM of Bao et al. \cite{Bao2017Simulation} with gravity of fluid, theory of Wen et al. \cite{wen2020Numerical} without considering the transition stage, Eqs.$\,$(\ref{Eq:SC8})  and $\,$(\ref{Eq:SV2}) during the water entry in freefall motion.}
  \label{Fig:freefall_motion_30}
\end{figure}

The case in \figurename$\,$\ref{Fig:CFD_HBF_Wu_comp} is also used for a validation for the present model. \figurename$\,$\ref{Fig:freefall_motion_30} shows the comparisons of the predictions of accelerations of wedge with $\beta=30^\circ$ between the CFD without gravity of fluid, BEM of Bao et al. \cite{Bao2017Simulation} with gravity of fluid, theory of Wen et al. \cite{wen2020Numerical} without considering the transition stage, Eq.$\,$(\ref{Eq:SC8}) and Eq.$\,$(\ref{Eq:SV2}) during the water entry in freefall motion. The agreement between the CFD result and the present model with acceleration effect, e.g., Eq.$\,$(\ref{Eq:SV2}), is good in both slamming and transition stages, while the theory of approximate solution without considering the transition stage \cite{wen2020Numerical} can only address the slamming stage. The present model without acceleration effect, e.g., Eq.$\,$(\ref{Eq:SC8}),  also shows large differences compared with the CFD result. The gravity effect has small influence on the hydrodynamic force in the transition stage for the cases of $Fr=1.9587$. For the high speed impact of higher $Fr$, the gravity of fluid can be neglected and the present model will have better predictions.

\begin{figure}
  \centerline{\includegraphics[width=9cm,height=6cm]{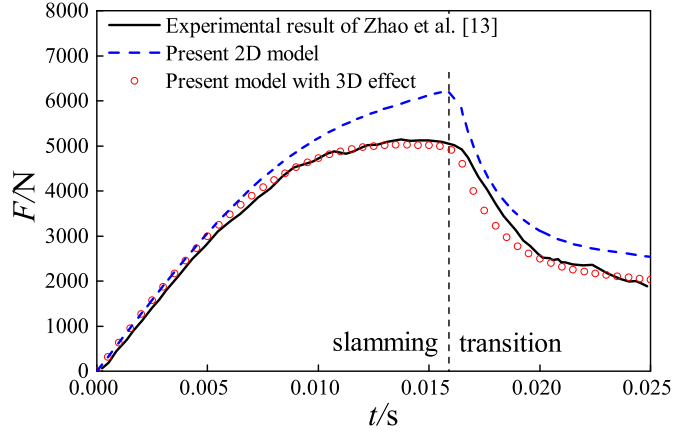}}
  \caption{The comparisons between the present predictions of forces and the experimental result of Zhao et al. \cite{zhao1996water} for the water entry of a wedge of $\beta=30^\circ$. The vertical velocity of the wedge is given by the experimental result of an optical sensor. The force coefficient of 3D effect is approximately formulated as a parabolic $f_{\rm 3D}=1-0.80(c/l)^2$ by using Meyerhoff's results \cite{meyerhoff1970added}.}
  \label{Fig:zhao_exp_30}
\end{figure}

A comparison of the forces between an experimental test of Zhao et al. \cite{zhao1996water} and the present model is shown in \figurename$\,$\ref{Fig:zhao_exp_30}. The case is a three-dimensional water entry with deadrise angle of $\beta=30^\circ$. The half width of wedge is $l=0.25\,\rm m$ and the thickness is $L=1\,\rm m$. The vertical speed of the wedge is given by the experimental result of an optical sensor. The acceleration of wedge is given by the derivative of vertical speed. The present 2D model has large discrepancies compared with the experimental result. As Zhao et al. indicated, the reason is due to the three dimensional (3D) effect and the force should be corrected by a force coefficient $f_{\rm 3D}$ (a function of $c/L$). They used Meyerhoff's results \cite{meyerhoff1970added} of $f_{\rm 3D}(0.25)=0.95$, $f_{\rm 3D}(0.4)=0.87$, $f_{\rm 3D}(0.5)=0.8$ from the added masses of thin rectangular plates with a generalization of Wagner theory to formulate the 3D effect. In this paper, a parabolic $f_{\rm 3D}=1-0.80(c/L)^2$ is approximately formulated by fitting from the above results. After using the 3D correction, the present model with 3D effect can predict the force of the experimental test.

To sum up, Eq.$\,$(\ref{Eq:SV2}) together with Eqs.$\,$(\ref{Eq:SC1}), (\ref{Eq:SC2}), Eq.$\,$(\ref{Eq:SV1}) and (\ref{Eq:total_constant}) can be used for both the predictions of hydrodynamic force in slamming and transition stages for the water entry of wedges with different deadrise angles in constant and varying speeds. 

\subsection{Comparisons with other theories}\label{Part:5.4}

For the slamming stage, the traditional added mass methods have the forms of added mass and the hydrodynamic force (half model)
\begin{equation}\label{Eq:SV3}
m_0 = \frac{1}{2}\rho {c^2}{C_a},
\end{equation}
\begin{equation}\label{Eq:SV4}
F = \frac{{{\rm{d}}({m_0}{U_w})}}{{{\rm{d}}t}} = \frac{1}{2}\rho U_w^2h{\cot ^2}\beta {C_{{\rm{const}}}} + \frac{1}{2}\rho {c^2}{a_w}{C_a},
\end{equation}
where $c=\gamma h\cot\beta$ for the traditional methods. \tablename$\,$\ref{Tab:gamma_and_Ca} shows the correction factor $\gamma$ of the wetted length, the dimensionless coefficient of constant speed impact $C_{\rm const}$  and the added mass coefficient $C_a$ of slamming stage for different theories \cite{von1929impact,wagner1932stoss,faltinsen1993sea,mei1999water,Tassin2014On}. The Wagner's new model \cite{wagner1932stoss} was proposed to improve the predictions of forces of constant speed impact for different deadrise angles, but the acceleration term remained unchanged. $\gamma_{\rm Mei}$ can be found in Refs.$\,$ \cite{mei1999water,wen2020Numerical}. $B_{2\rm Mei}$ and $A_{2\rm Mei}$ are calculated by direct pressure and velocity potential integrations of Mei et al.'s model \cite{mei1999water}. $C_{\rm Korobkin}$ is from the theoretical results of Korobkin  \cite{korobkin2004analytical}. $B_2$ and $A_2$ of the present model were calculated from the similarity solution \cite{dobrovol1969some}, and the results are shown in Ref.$\,$\cite{wen2020Numerical}. 
\begin{comment}
Since the numerical results of $k_1$ are only provided in the range of $\beta\in [15^\circ,\,\, 45^\circ]$ as shown in \figurename$\,$\ref{Fig:maximum_Cs_h0_BEM_CFD} ($b$), the $k_1$ of $\beta<15^\circ$ for the present model is given as $2/\pi$ from the Wagner theory as the deadrise angle is small. 
\end{comment}
\begin{table}
\centering
\begin{tabular}{cccc}
\toprule
Model&$\gamma$&$C_{\rm const}$&$C_a$\\
\midrule
Von Karman \cite{von1929impact}& 1&$\pi$&1\\
Wagner's original model \cite{wagner1932stoss}&$\frac{\pi}{{2}}$&$\frac{\pi^3}{4}$&$\frac{\pi}{{2}}$\\
Wagner's new model \cite{wagner1932stoss}&$\frac{\pi}{{2}}$&$\pi {\left( {\frac{\pi }{{2\beta }} - 1} \right)^2}\tan^2\beta$&$\frac{\pi}{{2}}$\\
Faltinsen \cite{faltinsen1993sea}&$\frac{\pi}{{2}}$&$\frac{\pi^3}{4}(1-\frac{\beta}{{2\pi}})^2$&$\frac{\pi}{2}(1-\frac{\beta}{{2\pi}})^2$\\
Mei et al. \cite{mei1999water}&$\gamma_{\rm Mei}$&$B_{2\rm Mei}\tan^2\beta$&$A_{2\rm Mei}(\tan\beta/\gamma_{\rm Mei})^2$\\
Tassin et al. \cite{Tassin2014On}&$\frac{\pi}{{2}}$&$C_{\rm Korobkin}$&$\frac{\pi }{2} - \left( {\frac{4}{\pi } - 1} \right)\tan \beta $\\
Present model&$\frac{1}{k_1}$&$B_2\tan^2\beta$&$A_2(k_1\tan\beta)^2$\\
\bottomrule
\end{tabular}
\caption{The correction factor $\gamma$ of the wetted length, the dimensionless coefficient of constant speed impact $C_{\rm const}$  and the added mass coefficient $C_a$ of slamming stage for different theories.}
\label{Tab:gamma_and_Ca}
\end{table}

\begin{figure}
  \centerline{\includegraphics[width=15cm,height=5cm]{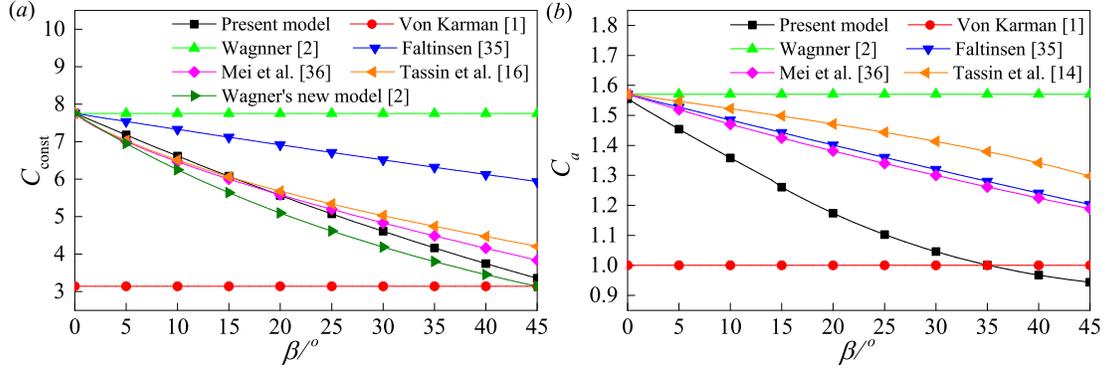}}
  \caption{The comparisons of the dimensionless coefficient of ($a$) constant speed impact $C_{\rm const}$   and ($b$) added mass coefficient $C_a$   of the slamming stage between the present model and other theories \cite{von1929impact, wagner1932stoss, faltinsen1993sea,mei1999water,Tassin2014On}. }
  \label{Fig:Ca_comparision}
\end{figure}

In early researches of Refs.$\,$\cite{von1929impact,wagner1932stoss,faltinsen1993sea,mei1999water,korobkin2004analytical} for the slamming stage,  the hydrodynamic force of constant speed impact received the most attention and the researchers proposed their models by comparing the results of similarity solution \cite{dobrovol1969some,zhao1993water} to correct the underestimated results of Von Karman \cite{von1929impact} and overestimated results of Wagner's original model \cite{wagner1932stoss}, as shown in \figurename$\,$\ref{Fig:Ca_comparision} ($a$).  The models of Mei et al. \cite{mei1999water}, Tassin et al. \cite{Tassin2014On} and Wagner's new model \cite{wagner1932stoss} have been approximately close to the results of similarity solution, but these models actually violate the framework of added mass methods. Their $C_{\rm const}$ are no longer $2C_a\gamma^2$, which are different from the models of Von Karman \cite{von1929impact}, Wagner \cite{ wagner1932stoss} (original model) and Faltinsen \cite{faltinsen1993sea}. Although these models \cite{mei1999water,Tassin2014On,wagner1932stoss} brought good predictions for the constant speed impact, they still can not properly model the acceleration effect, as shown in \figurename$\,$\ref{Fig:Ca_comparision} ($b$). The Faltinsen's model cannot properly address both the constant speed impact and acceleration effect. The present model formulates the models of constant speed impact and acceleration effect directly from the approximate solution \cite{wen2020Numerical} based on the similarity solution of Dobrovol'skaya \cite{dobrovol1969some}, regardless of the framework of added mass methods and the consistency of $C_{\rm const}$ and $C_a$. The validations by numerical results in Ref.$\,$\cite{wen2020Numerical} and Figs.$\,$\ref{Fig:predictions_30deg_CasesABC} and \ref{Fig:predictions_14_20_45_CasesDEF} have completely verified the present model.

For the transition stage, Tassin et al.'s model \cite{Tassin2014On}, the MWM of Wen et al.  \cite{wen2022Modified} and the present model can address both the slamming and transition stages, while the models of Ref.$\,$\cite{von1929impact,wagner1932stoss,faltinsen1993sea,mei1999water} can only work in the slamming stage. Wen et al. had greatly improved the predictions of hydrodynamic forces for the constant speed impact in the range of $\beta\in[15^\circ,\,\,35^\circ]$ compared with the linear FBC of Tassin et al.  and the improved model of zero pressure condition \cite{logvinovich1969hydrodynamics,Tassin2014On}. The accuracy of present model is close to Wen et al.'s MWM, but with a much larger range of deadrise angle. Besides, the acceleration effect in the transition stage is missing in MWM. Tassin et al.'s model provided the formulation of acceleration effect and the $C_a$ is given by Eq.$\,$(\ref{Eq:Ca_Tassin_trans}). The comparison of $C_a$ of $\beta=30^\circ$ between the present model and Tassin et al.'s model is shown in \figurename$\,$\ref{Fig:Ca_comparision} ($b$). In contrast to an increase of $C_a$ in the transition stage, the $C_a$ of Tassin et al.'s model declines and finally reaches a constant value lower than $C_{a0}$, which is inconsistent with the HBF results in \figurename$\,$\ref{Fig:Ca_comparision} ($b$).

In general, we rewrite the formula of hydrodynamic force and extend it to the transition stage for the constant and varying speed impacts. Although the present model is more sophisticated, it works better than the traditional added mass methods and the asympotic theories and can work in a larger range of deadrise angle. 

\section{Conclusions}
In this paper, the transition stage of the water entry of a linear wedge with constant and varying speeds is numerically studied by FVM with VOF and HBF. The hydrodynamic force acting on the wedge is formulated by a combination of the numerical results and theoretical results of the steady supercavitating flow. The fluid is assumed to be incompressible, inviscid, weightless and with negligible surface tension, and the flow is irrotational for the high speed impact. For the constant speed impact, the similitude of the slamming coefficients of different deadrise angles in the transition stage is found by scaling the difference between the maximum values in the slamming stage and the results of steady supercavitating flow. The formulation of the hydrodynamic force is conducted based on the similitude of the declining forces in the transition stage together with the linear increasing results in the slamming stage. For the varying speed impact, the acceleration effects on the pressure distribution and force acting on the wedge surface are revealed. The hydrodynamic force caused by the acceleration effect in the transition stage is formulated by an added mass coefficient with an averaged increase of $27.13\%$ compared with that of slamming stage. A general expression of the slamming coefficient with the deadrise angles from $5^\circ$ to $70^\circ$ in both the slamming and transition stages is thus proposed for the constant and varying speed impacts, and its predictions are in good agreement with the numerical and experiment results. Thus, it can be treated as a 2D water entry model the 2.5D method being used on the take-off and water landing of seaplanes, and for the strip theory or 2D+t theory being used on the hull slamming.

\begin{comment}
The present semi-analytical solutions is still limited to the water entry of linear wedges, corresponding to the cross-section A and C in \figurename$\,$\ref{Fig:seaplane_description}, the solution for the water entry of curved wedges corresponding to cross-section B in \figurename$\,$\ref{Fig:seaplane_description} will be addressed in our further work.
\end{comment}

\section*{Acknowledgements}

This work was partially supported by the National Natural Science Foundation of China (No. 12072014 and No. 11721202).

\section{Appendix}\label{App}
\subsection{The theory of steady supercavitating flow}\label{App:1}
\begin{figure}
  \centerline{\includegraphics[width=10cm,height=5cm]{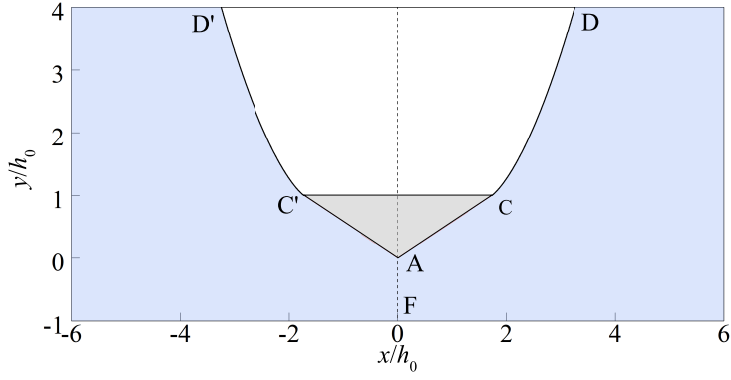}}
  \caption{Free surface of the steady supercavitating flow for a 2D wedge of $\beta=30^\circ$.}
  \label{Fig:infinite_stage_free_surface}
\end{figure}

\begin{figure}
  \centerline{\includegraphics[width=12cm,height=2cm]{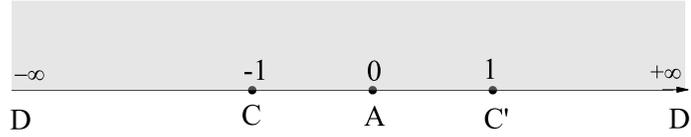}}
  \caption{ The $\tau$ plane of the steady supercavitating flow.}
  \label{Fig:geometry_mapping}
\end{figure}

\begin{figure}
  \centerline{\includegraphics[width=9cm,height=3.9cm]{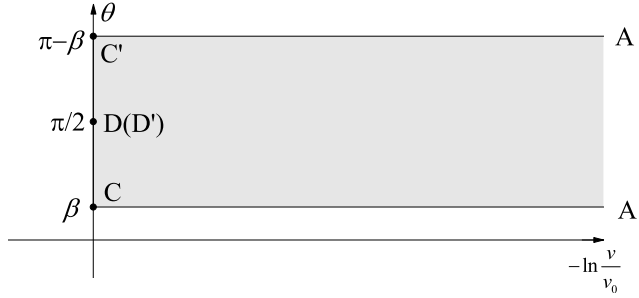}}
  \caption{The $\omega$ plane of the steady supercavitating flow.}
  \label{Fig:velocity_mapping}
\end{figure}

The water entry in the infinite penetration depth becomes a steady supercavitating flow with an incoming currency with velocity $v_0$. The theory can be extended from the water entry on a plate by the theory of Gurevich \cite{gurevich1965theory}. The free surface CD and $\rm C'D'$ of impact flow is shown in \figurename$\,$\ref{Fig:infinite_stage_free_surface} after the solution is solved. The direction of incoming flow is upward and the stream-function on the streamline passing A is 0. The velocity potential at A is chosen as 0. A parameter plane $\tau$ mapping to the geometry in \figurename$\,$\ref{Fig:infinite_stage_free_surface} is shown in \figurename$\,$\ref{Fig:geometry_mapping} and thus the complex velocity potential of impact flow can be calculated as
\begin{equation}\label{Eq:App1}
w=\phi_0\tau^2,
\end{equation}
where $\phi_0$ is the velocity potential at C and $C'$. Another parameter plane $\omega=v_0\frac{{\rm d}z}{{\rm d}w}=v_0e^{{\rm i }\theta}/v$ ($\rm i$ being the imaginary unit) mapping to the velocity is shown in \figurename$\,$\ref{Fig:velocity_mapping} and can be formulated by the Schwarz-Christoffel formula as
\begin{equation}\label{Eq:App2}
\omega \left( \tau \right) = \left( {1 - \frac{{2\beta }}{\pi }} \right)\ln \left( {\sqrt {1 - 1/{\tau^2}}  + \frac{{\rm{i}}}{\tau}} \right) + {\rm{i}}\frac{\pi }{2}.
\end{equation}
The complex coordinate can be derived
\begin{equation}\label{Eq:App3}
z = \int {\frac{{{\rm{d}}z}}{{{\rm{d}}w}}} \frac{{{\rm{d}}w}}{{{\rm{d}}\tau}}{\rm{d}}\tau = \frac{{2{\phi _0}}}{{{v_0}}}{\rm{i}}\int {{{\left( {\sqrt {1 - 1/{\tau^2}}  + \frac{{\rm{i}}}{\tau}} \right)}^{1 - \frac{{2\beta }}{\pi }}}} \tau{\rm{d}}\tau.
\end{equation}
By considering the distance from A to C, the velocity potential at C and $C'$ can be determined
\begin{equation}\label{Eq:App4}
\frac{{2{\phi _0}}}{{{v_0}}} = \frac{l}{{A\cos \beta }},
\end{equation}
where 
\begin{equation}\label{Eq:App5}
A = {\int_0^1 {\left( {\frac{{1 + \sqrt {1 - {\tau ^2}} }}{\tau }} \right)} ^{1 - \frac{{2\beta }}{\pi }}}\tau {\rm{d}}\tau 
\end{equation}
Therefore, the $x$ on AC is given as
\begin{equation}\label{Eq:App6}
x\left( \tau  \right) =   \frac{l}{A}{\int_0^\tau  {\left( {\frac{{1 + \sqrt {1 - {\tau ^2}} }}{\tau }} \right)} ^{1 - \frac{{2\beta }}{\pi }}}\tau {\rm{d}}\tau,
\end{equation}
and the velocity on AC is
\begin{equation}\label{Eq:App7}
v = {v_0}{\left( {\frac{{1 - \sqrt {1 - {\tau ^2}} }}{\tau }} \right)^{1 - \frac{{2\beta }}{\pi }}}
\end{equation}
The pressure coefficient $C_p$ has the form
\begin{equation}\label{Eq:App8}
{C_p} = 1 - {\left( {\frac{{1 - \sqrt {1 - {\tau ^2}} }}{\tau }} \right)^{2\left( {1 - \frac{{2\beta }}{\pi }} \right)}}.
\end{equation}
\begin{figure}
  \centerline{\includegraphics[width=9cm,height=5.4cm]{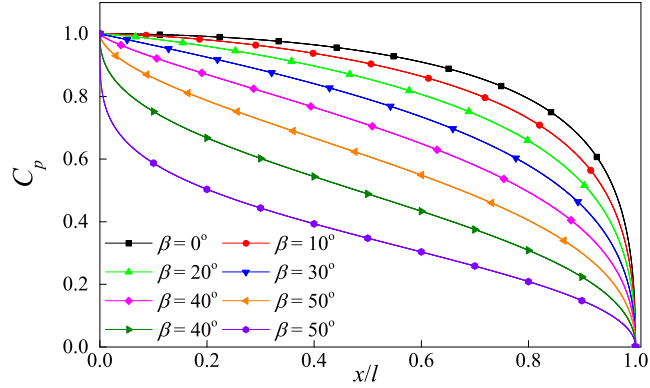}}
  \caption{The $C_p$ distributions of different deadrise angles for the steady supercavitating flow.}
  \label{Fig:Cp_infinite_stage}
\end{figure}
The $C_p$ distributions of different deadrise angles $\beta$ are shown in \figurename$\,$\ref{Fig:Cp_infinite_stage}. The $C_p$ is 1 at the stagnation A and 0 at the separation  C for a steady supercavitating flow. The free surface profile in \figurename$\,$\ref{Fig:infinite_stage_free_surface}  is given by the following $x$ and $y$ expressions
\begin{equation}\label{Eq:App9}
x\left( \tau  \right) =   l - \frac{l}{{A\cos \beta }}{\rm{Re}}\left[ {{\rm{i}}{{\int_1^\tau  {\left( {\sqrt {1 - 1/{\tau ^2}}  + \frac{{\rm{i}}}{\tau }} \right)} }^{^{1 - \frac{{2\beta }}{\pi }}}}\tau {\rm{d}}\tau } \right],
\end{equation}
\begin{equation}\label{Eq:App10}
y\left( \tau  \right) = l\tan \beta  + \frac{l}{{A\cos \beta }}{\rm{Im}}\left[ {{\rm{i}}{{\int_1^\tau  {\left( {\sqrt {1 - 1/{\tau ^2}}  + \frac{{\rm{i}}}{\tau }} \right)} }^{^{1 - \frac{{2\beta }}{\pi }}}}\tau {\rm{d}}\tau } \right].
\end{equation}

\begin{figure}
  \centerline{\includegraphics[width=9cm,height=5.4cm]{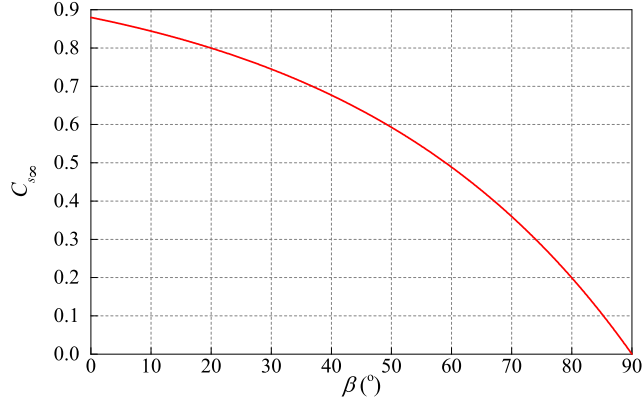}}
  \caption{The slamming coefficient $C_{s\infty}$ of the steady supercavitating flow.}
  \label{Fig:Cs0_infinite_stage}
\end{figure}

The force acting on the wedges is calculated by integrating the pressure with the $C_p$ expression Eq.$\,$(\ref{Eq:App8}). The slamming coefficient $C_{s\infty}$ of different deadrise angles is shown in \figurename$\,$\ref{Fig:Cs0_infinite_stage}. The result of $\beta=0$ is equal to 0.88, which is consistent with that of water entry on a plate derived by Gurevich \cite{gurevich1965theory}. The $C_{s\infty}$ decreases  with the increasing deadrise angle and reaches 0 at $\beta=90^\circ$.

Korvin-Kroukovsky and Chabrow \cite{korvin1948discontinuous} in 1948 had already presented the calculation of pressure distribution on the wedge surface. The present theory is consistent with their model.

\section*{References}

\bibliography{elsarticle-template}

\end{document}